\documentclass[fleqn,usenatbib,useAMS]{mnras}

\usepackage{graphicx}	
\usepackage{epstopdf}
\usepackage{amsmath}	
\usepackage{amssymb}	
\usepackage{multicol}        
\usepackage{bm}		
\usepackage{pdflscape}	

\usepackage{colortbl}  
\usepackage{xcolor}
\usepackage{array}
\usepackage{verbatim}

\usepackage[T1]{fontenc}

\usepackage{threeparttable} 

\title{Four Changing-Look Active Galactic Nuclei Found From Optical Variations}

\author[Zhu et al.]{
	Li-Tao Zhu$^{1}$, 
	Jie Li$^{2,3}$,
	Zhongxiang Wang$^{1,4,}$\thanks{E-mail: wangzx20@ynu.edu.cn}
	Ju-Jia Zhang$^{5,6}$
	\\
	$^{1}$Department of Astronomy, School of Physics and Astronomy, Yunnan University, Kunming 650091, China; wangzx20@ynu.edu.cn
	\\
       $^{2}$Laboratory for Research in Galaxies and Cosmology, Department of 
Astronomy, University of Science and Technology of China, \\
Hefei 230036, China\\
	$^{3}$School of Astronomy and Space Science,
University of Science and Technology of China, Hefei 230026, China\\
    $^{4}$Shanghai Astronomical Observatory, Chinese Academy of Sciences, Shanghai 200030, China\\
    $^{5}$Yunnan Observatories, Chinese Academy of Sciences, Kunming 650216, 
    China\\
    $^{6}$Key Laboratory for the Structure and Evolution of Celestial Objects, 
    Chinese Academy of Sciences, Kunming 650216, China\\
}

\date{Accepted XXX. Received YYY; in original form ZZZ}

\pubyear{2023}

\begin{document}
	\label{firstpage}
	\pagerange{\pageref{firstpage}--\pageref{lastpage}}
	\maketitle
	
	\begin{abstract}
	We report the finding of four changing-look (CL) active galactic 
		nuclei (AGN). We selected these sources due to their
potential as interesting targets when considering their relatively-large 
optical flux variations and related mid-infrared 
		flux variations. To identify their 
CL feature, we use archival spectra from the Sloan Digital Sky Survey (SDSS) 
	taken at least 8 years ago as well as spectra taken recently from 
	the Transient Name Server (TNS) and with the 2.4-m LiJiang 
	telescope (LJT).  We study
the sources' spectral changes by fitting and determining 
the H$_\alpha$ and H$_\beta$ components and verify their CL behavior. 
When comparing the TNS and/or LJT spectra to the SDSS ones, all 
	four sources showed the appearance
of a broad or a stronger broad H$_\alpha$ component and a relatively weak
		broad H$_\beta$ component. As two of the four sources 
		are established to have a brighter-and-bluer feature in the 
		photometric data, during the time periods in which the TNS
		and LJT spectra were taken, this feature likely accompanied
		the turn-on of the broad components. Thus, we suggest that
		this brighter-and-bluer feature can be used as
		a criterion for efficiently finding CL sources among
		previously spectroscopically classified type 2 AGN,
		such as from among the sources provided by the SDSS.
	\end{abstract}
	
	\begin{keywords}
		galaxies: active --- quasars: emission lines
	\end{keywords}
	
	
	
	
\section{INTRODUCTION}
	
According to the framework of the unification scheme for Active Galactic 
Nuclei (AGN), they can be classified into two types, type 1 and type 2, 
depending on our viewing angle of the galactic central region that contains
the power source, an actively accreting supermassive black hole (SMBHs; e.g.,
\citealt{law87, ant93, up95, tad08}).
For type 1, 
broad ($\gtrsim 1000 \mathrm{~km} \mathrm{~s}^{-1}$) and 
narrow ($\leqslant 1000 \mathrm{~km} \mathrm{~s}^{-1}$) 
emission lines are seen in optical or ultraviolet (UV) spectra as we are viewing
the centers with low inclination angles, while for type 2,
only narrow lines can be seen since the broad line regions (BLRs), 
presumably closer
to the centers than the narrow line regions (NLRs), are obscured by 
the dusty tori that surround the SMBHs. This unification scheme has worked
well in understanding of various AGN phenomena but is now being
challenged by
the so-called Changing-Look AGN (CLAGN) discovered in recent years.

CLAGN can be identified from either X-ray or optical/UV spectroscopy. 
In X-rays, the majority of the CLAGN can be seen to have undergone 
transitions between Compton-thick and Compton-thin states
(e.g., \citealt{mgm03, pfr+07, bpc+09, rye09, mbd+12}; see also \citealt{rt22}
and references therein). The transitions, intricately linked to the levels
of obscuration, are possibly attributed to various causes, 
such as
eclipses, fluctuations in the ionization state of the obscuring gas, 
outflows, or the switching on-and-off of the central engine.

In optical (and UV) wavelengths, occasional cases of striking appearances 
or disappearances of broad emission lines (BELs), in particular a pronounced 
$H_\beta$ component, have been reported in early years 
(e.g., \citealt{to76, crp+86, sbw93, ajk+99, eh01, ddc+14, spg+14}). This 
includes the first quasar case \citep{lcm+15}. Because of the intriguing 
features and their importance in helping further reveal the physical processes
in AGN, 
systematic searches for CLAGN have recently been conducted given the advent of
rich data from different surveys. As a result, an increasing 
number of CLAGN have been reported
(e.g., \citealt{rac+16, mrl+16, ywf+18, swj+20, gpa+22}). 

The methods performed in these searches were to compare the spectra from
different epochs all the while
taking into consideration flux variations, since
spectral changes were notably accompanied with large
flux variations (see, e.g., \citealt{lop+23} and references therein). 
We have been working on studies of $\sim$40
AGN flux-variation cases, selected mainly from the optical imaging 
survey data from the Zwicky Transient Facility (ZTF; \citealt{bkg+19}),
and have found them showing different interesting patterns 
(e.g., \citealt*{lwz23}).
Most of these AGN have archival spectra from the Sloan Digital Sky Survey
(SDSS; \citealt{AAA+09}), and by comparing that with the spectra reported 
at the Transient Name Server (TNS) and/or spectra obtained by us with the 
2.4-m LiJiang Telescope (LJT; \citealt{wan+19}), 
we were able to identify four CLAGN among them. They are named as J0113+0135, 
J1127+3546, J1223+0645, and J1513+1759 (hereafter J0113, J1127, J1223, 
and J1513, respectively).
Information for them is provided in Table~\ref{tab:src}.
We note also that one of them, J0113, has already been reported 
by \citet{lmb+22} as a CLAGN.

In this paper, we report the results that lead to the identification of the 
four CLAGN.  
In Section~\ref{sec:obs}, we describe the optical and infrared (IR) 
photometric and spectroscopic data we used in the studies;
as we conducted spectroscopic observations ourselves for three of them, 
we also 
provide the details of our observations and data reductions.
The spectrum results and analysis, which show the changing-look (CL) feature
of the four sources, are presented in Section~\ref{sec:res}. Finally
we discuss
the results in Section~\ref{sec:dis}, where we also point out 
that a brightness-color behavior of AGN could be a key criterion for efficiently
selecting CLAGN candidates.
Throughout this paper, we adopted cosmological parameters
from the Planck mission \citep{paa+18}, 
with \emph{$H_0$} = 67 km\,s$^{-1}$\,Mpc$^{-1}$.

\section{Data and Observations}
\label{sec:obs}

\subsection{Archival Data}

In order to find AGN with significant flux variations,
we mainly used the ZTF $g$- and $r$-band ($zg$ and $zr$ respectively)
data. Magnitudes of the two bands are in the AB photometric system
\citep{bkg+19}, similar to those \citep{ztf2pan} provided in the Panoramic
Survey Telescope and Rapid Response System \citep{pan16}.
We required catflags = 0 and $chi < 4$ when querying ZTF data to be
able to construct a clean and good-quality light curve for a source,
and when 
interesting sources were found, we added other available photometric
data in order to obtain as complete a light curve
as was possible at multiple bands.
The other data included were cyan (420--650\,nm; $ac$) and orange 
(560--820\,nm; $ao$) bands, where the two cover the SDSS
filters $g+r$ and $r+i$ wavelength ranges respectively,
from the Asteroid Terrestrial-impact Last Alert System (ATLAS; 
\citealt{tdh+18}), $V$ band (Vega based system) 
from the Catalina Real-Time Transient 
Survey (CRTS; \citealt{ddm+09}),
and mid-IR (MIR) W1 (3.4\,$\mu$m) and W2 (4.6\,$\mu$m) bands from the Post-Cryo
survey of the Wide field Infrared Survey Explorer (WISE; \citealt{wri+10}).

All four sources have an optical spectrum from the SDSS database, while
three of the four have an spectrum each from the TNS database; 
J0113 was reported by \citet{pda+22},
J1127 by \citet{jf21}, and J1513 by \citet{hin22}.
We also obtained spectra ourselves for three of the AGN for purposes
of either confirmation or 
identification; the spectroscopic observation details are described in 
the following Section~\ref{subsec:ljt}.

\begin{table}
	\centering
	\caption{Source information for the four CLAGN}
	\label{tab:src}
	\begin{tabular}{lcccc}
\hline
		Source & R.A. (J2000)  & Decl. (J2000)  & $z$  & Mag$^a$ \\ 
		\hline
		J0113 & $01^h$$13^m$$11^s.8$  & +$01^\circ$$35^\prime$$42^{\prime \prime}$.5 & 0.24 & 19.23   \\
		J1127 & $11^h$$27^m$$59^s.3$ & +$35^\circ$$46^\prime$$44^{\prime \prime}$.8 & 0.075 & 17.78   \\
		J1223 & $12^h$$23^m$$10^s.6$ & +$06^\circ$$45^\prime$$54^{\prime \prime}$.9  & 0.24 & 17.99  \\
		J1513 & $15^h$$13^m$$20^s.9$ & +$17^\circ$$59^\prime$$16^{\prime \prime}$.1 & 0.13 & 17.01  \\
		\hline
	\end{tabular}\\
	\begin{tablenotes}[]
	\item $^a$ Magnitudes were ZTF $g$-band values at the times of the most recent spectroscopic observations, except for J1513+1759 whose ATLAS $ao$ value is given.
	\end{tablenotes}            
\end{table}      

\subsection{LJT observations}
\label{subsec:ljt}

Using the LJT, we conducted spectroscopic observations of three of the sources.
The Yunnan Faint Object Spectrograph and Camera (YFOSC),
whose detector is a 2048$\times$4096\,pixel$^2$ back-illuminated 
CCD with a pixel scale of 0.283$''$ pixel$^{-1}$, was used for spectroscopy.
In all exposures,
a G3 grism was chosen, providing a wavelength coverage of 340--910\,nm and
a spectral dispersion of 0.29\,nm\,pixel$^{-1}$. A long slit with a
width of 2.5$''$ was chosen for all exposures. Along with each of
the exposures of the sources, spectra of a He-Ne lamp and a 
spectrophotometric standard were also taken for the wavelength and
flux calibration, respectively. Information for the observations,
which includes the date, seeing, exposure time, and standard star, 
are given in Table~\ref{tab:obs}.

Using the IRAF tasks, the spectrum-image data were processed by performing
bias-subtraction and flat-fielding. Spectra of the sources were then
extracted, with wavelength and flux calibrations conducted.

        \begin{table}
        	\centering
        	\caption{Information for spectroscopic observations with LJT}
        	\label{tab:obs}
        	\begin{tabular}{lcccc}
        		\hline
        		Target & Date & Seeing & Exposure & Standard \\ 
 &  &  ($''$) & (sec) & \\
        		\hline
        		J0113 & 2022-10-20 & 0.75 & 2400 & BD+33d2642 \\
        		J1223 & 2022-04-23 & 1.46 & 1500 & Feige110 \\
        		J1513 & 2023-05-07 & 1.03 & 1800 & BD+33d2642 \\
        		\hline
        	\end{tabular}
        \end{table}  
                

\section{Analysis and Results}
\label{sec:res}
		
Visual inspection of the spectra of the four AGN taken at different epochs 
suggested the presence of the CL feature.
We employed the Python QSO fitting code (PyQSOFit; \citealt{gsw18})
to obtain measurements on the key emission lines, H$_{\alpha}$ and H$_{\beta}$
in the spectra.
From the spectral fitting, the full-width at half maximum (FWHM), equivalent 
width (EW), and line flux for each of the two lines were obtained. In addition,
the peak wavelength for each line's broad component was also obtained.
The results are given in Table~\ref{tab:line}, and
the fitting details are presented in Figures~\ref{fig:sf1}--\ref{fig:sf4} in 
Section Appendix~\ref{sec:app}.

However for the results, two caveats should be taken into consideration.
The first is the line broadening due to the seeing or
the slit width. Because the SDSS and TNS spectra are the archival data,
we only estimated the effect in our LJT observations. Since the seeing
was smaller than the slit width in the exposures, the broadening due to the
former approximately had FWHM$_{\rm seeing}\simeq 7.7$, 15.0, and 10.6\,\AA\ 
for the LJT spectra of J0113, J1223, and J1513 (cf., Table~\ref{tab:obs}), 
respectively. The corresponding values in units of kilometer
per second at the observed H$_{\beta}$ (H$_{\alpha}$)
line are approximately 380, 750, and 570 (280, 550, and 420), respectively.
These FWHM$_{\rm seeing}$ should be subtracted from FHWMs given in 
Table~\ref{tab:line} in quadrature, and they can reduce the FWHMs of the
narrow lines upto $\sim$23\%, while they obviously do not significantly
affect the FWHMs of the broad lines.

The second is the uncertainties on the spectra, which should be 
dominated
by systematic ones. To estimate them, we checked the fluxes of several 
continuum regions of a TNS or LJT spectrum that have a roughly same flux level
but different noise levels. The average flux difference with respect to that
of the continuum region of the lowest noise was considered as the systematic 
uncertainty of a spectrum. The uncertainties estimated this way are 
approximately 10\%, 9\%, and 
17\% for the LJT spectra of J0113, J1223, and J1513, respectively,
and 6\%, 3\%, and 9\% for the TNS spectra of J0113, 
J1127, and J1513, respectively. These uncertainties should be considered
as the systematic ones to the measurements of the line features given in 
Table~\ref{tab:line}.

\begin{figure*}
\centering
	\includegraphics[width=0.69\linewidth]{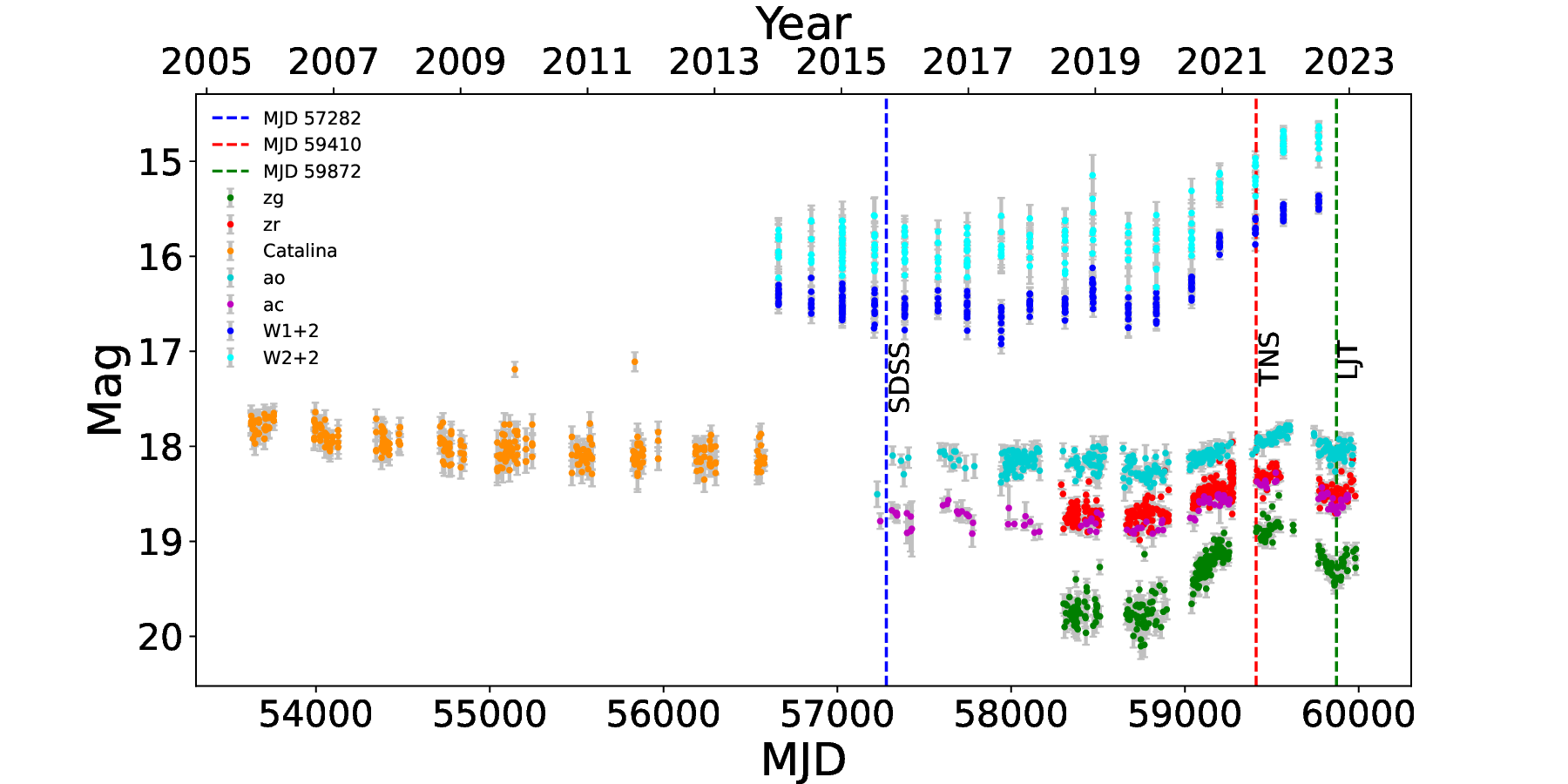}
	\includegraphics[width=0.69\linewidth]{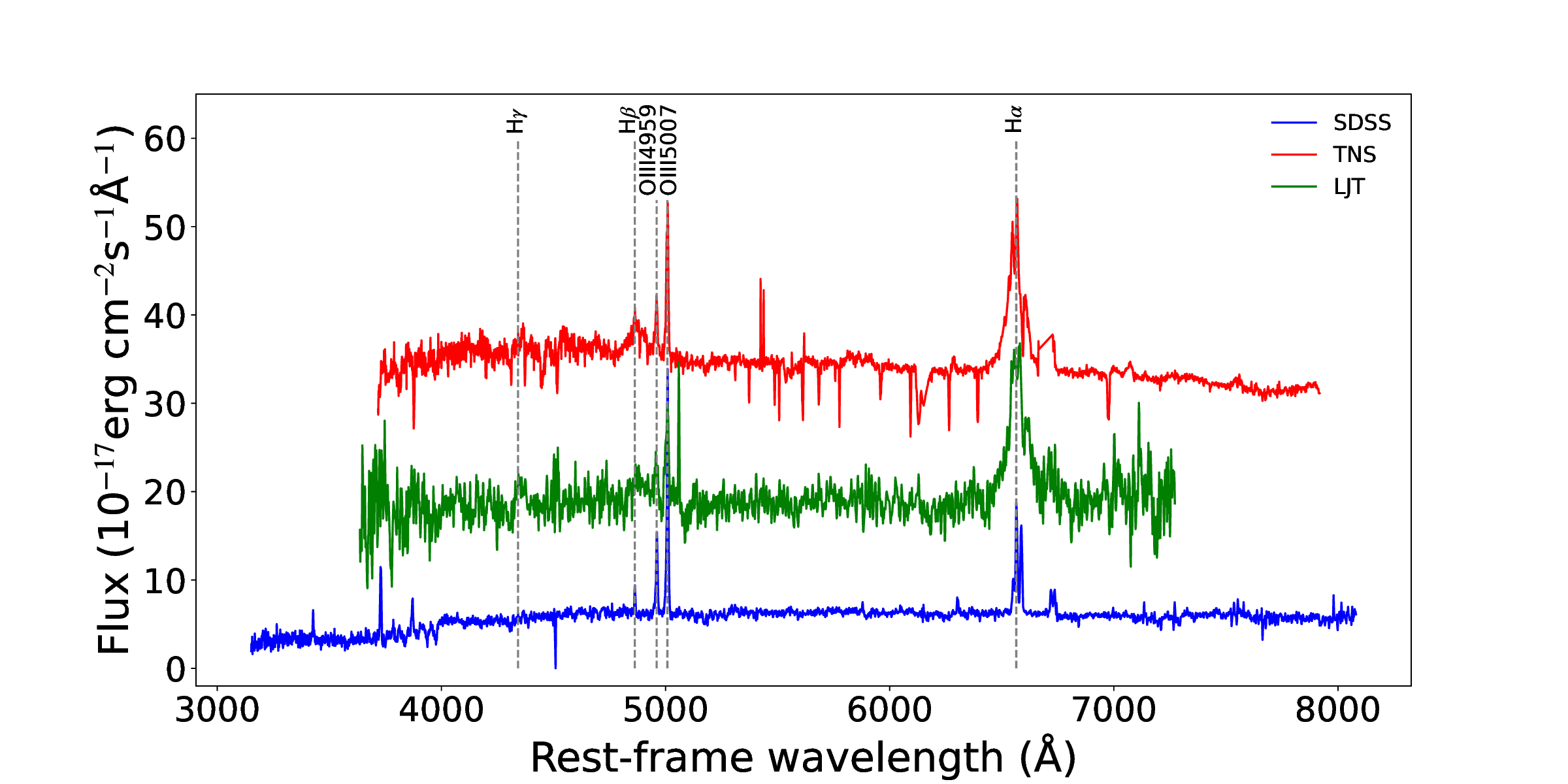}
		\caption{Optical and MIR light curves ({\it top} panel)
		and optical spectra ({\it bottom} panel) of J0113+0135. The 
		spectrum taking times are marked by the dashed lines in the
		top panel. In the bottom panel, the spectra are shifted 
		vertically for clarity, and the key emission lines are 
		indicated.}
			\label{fig:src1}
	\end{figure*}

\begin{figure}
\centering
\includegraphics[width=0.89\linewidth]{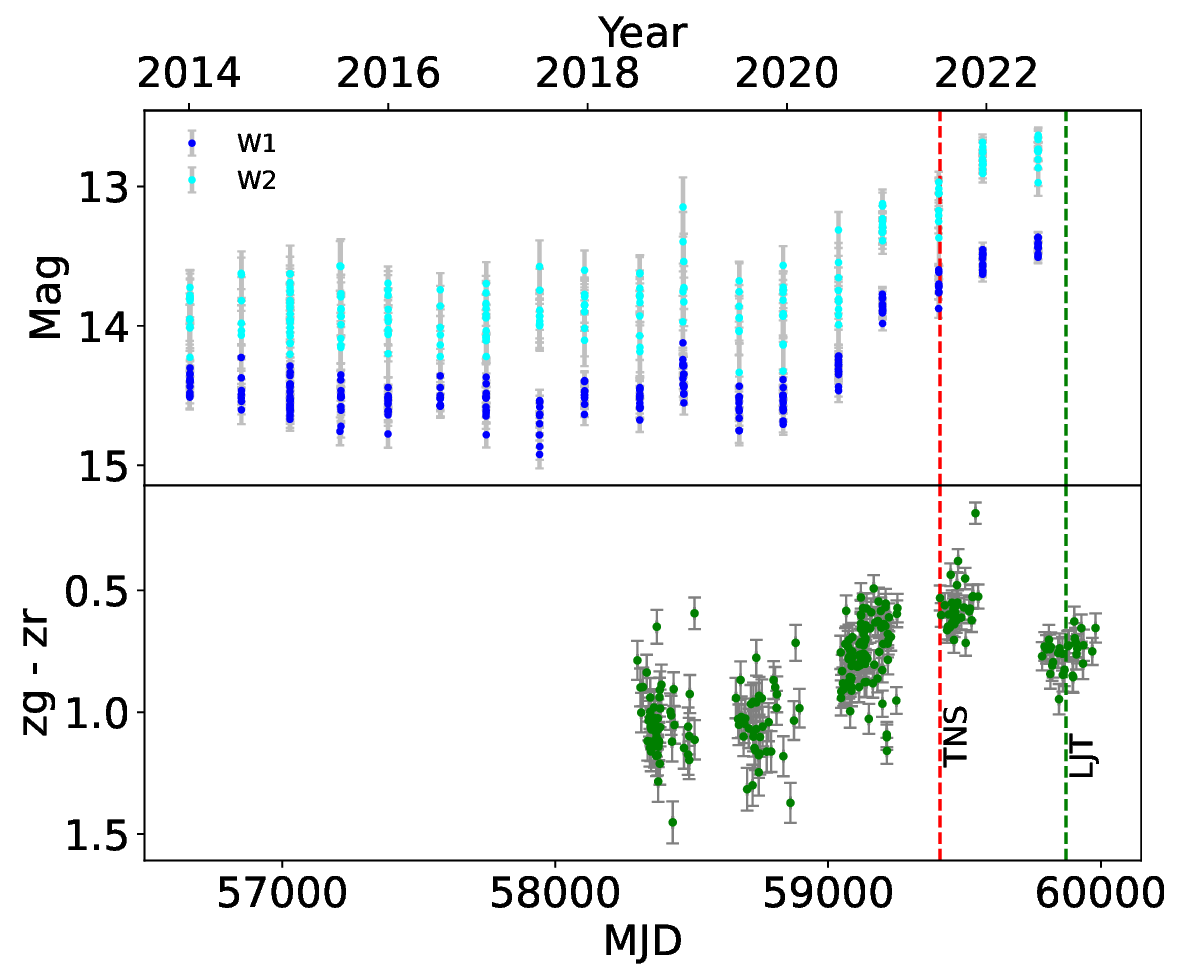}
	\caption{{\it Bottom}: $zg-zr$ color variations of J0113+0135. 
	The dashed lines mark the TNS and LJT spectra taking times. {\it Top}:
	Detailed view of the MIR W1 and W2 light curves of the source.
		\label{fig:c1}}
\end{figure}

\subsection{J0113+0135}
	
This source had a flat $V$-band light curve in the past,
but its ZTF light curves show a $\sim -$0.5\,mag rise after MJD~59000.
The MIR W1 and W2 light curves also show an accompanying rise, 
appearing to have lasted longer than the optical rise
(Figure~\ref{fig:src1}). 
Broad emission lines are clearly visible when comparing
the spectrum from the TNS (taken near the peak of the rise)
with the SDSS spectrum (taken on MJD~57282), and the 
broad-line feature was still present in our LJT spectrum (taken on MJD~59872),
when the optical light curves were already turning downwards.
The spectral fitting results indicate that the broad H$_{\alpha}$ and 
H$_{\beta}$ 
components had FWHMs of $\sim$5000--6000\,km\,s$^{-1}$. The redshift value was 
determined to be $\simeq$0.24. Following \citet{vp06}, the mass of the SMBH
$M_{\mathrm{BH}}$ was estimated from
        \begin{equation}
	\log\left(\frac{M_{\mathrm{BH}}}{M_{\odot}}\right) =\log\left[\left(\frac{\mathrm{FWHM}(\mathrm{H}_{\beta})}{\mathrm{km}\,\mathrm{s}^{-1}}\right)^{2}\left(\frac{L_{5100}}{10^{44}\,\mathrm{erg}\,\mathrm{s}^{-1}}\right)^{0.5}\right]+0.91,
        \end{equation}
where $L_{5100}$ is the luminosity at 5100\,\AA. The mass is 
$\sim$$10^{8.2}\ M_{\odot}$ based on the LJT measurement (Table~\ref{tab:line}).

As there was a significantly larger/faster flux rise in the blue $zg$ band,
clearly shown in Figure~\ref{fig:src1}, 
we further checked the details regarding it. A $zg-zr$ color diagram,
derived in approximation from the data points of $zg$ and $zr$ taken 
within one day, is shown in Figure~\ref{fig:c1}. 
Initially at $\sim$1.0\,mag, the color decreases, reaching to as low 
as $\sim$0.5\,mag before rising back to $\sim$0.75\,mag at the time when
the LJT spectrum was taken.
In addition, we found that the color 
changes follow a brighter-and-bluer pattern, $zg-zr\propto 0.59zr$. 
The magnitude-color diagram showing the derived relationship is presented in
Figure~\ref{fig:cm} in Appendix~\ref{sec:cm}.

Unfortunately, no spectra were taken
right before the color change at MJD~59000, so we do not know if the CL
feature occurred in tandem with the change or not. It is interesting to see
that the MIR
fluxes also increased, with magnitude changes 
of $\sim -$1\,mag in both W1 and W2 bands, and that it did not follow the 
flux fallback
seen in the optical variations. We did not conduct tests to determine the 
correlation behavior between
the optical and MIR bands, such as, checking if there were any time delays
in the MIR flux increases with respect to the optical rise, since the simple 
light-curve shapes (as well as the sparse data points in the MIR bands) 
would be unlikely to provide definitive results.

\begin{figure*}
			\includegraphics[width=0.69\linewidth]{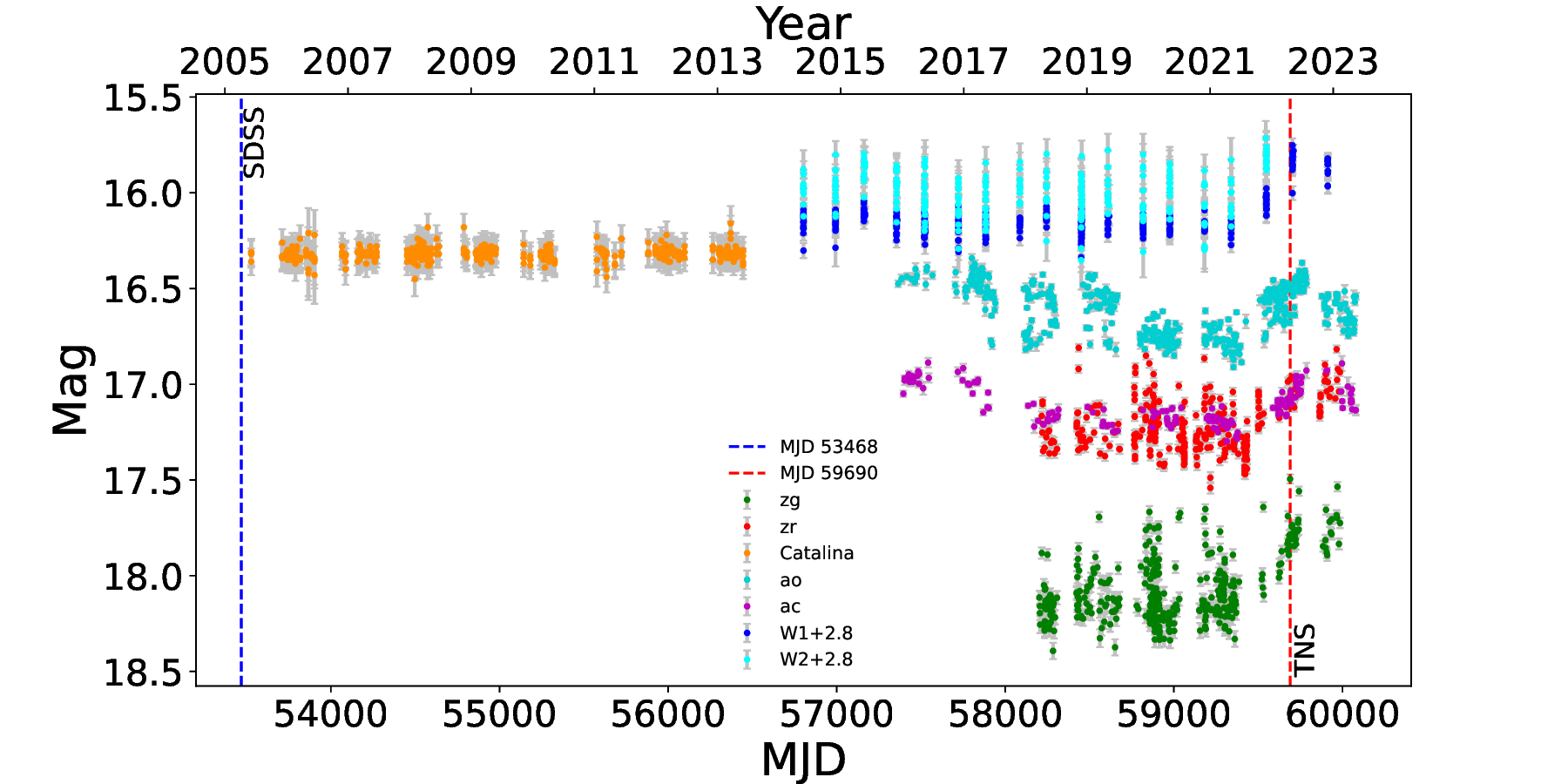}
			\includegraphics[width=0.69\linewidth]{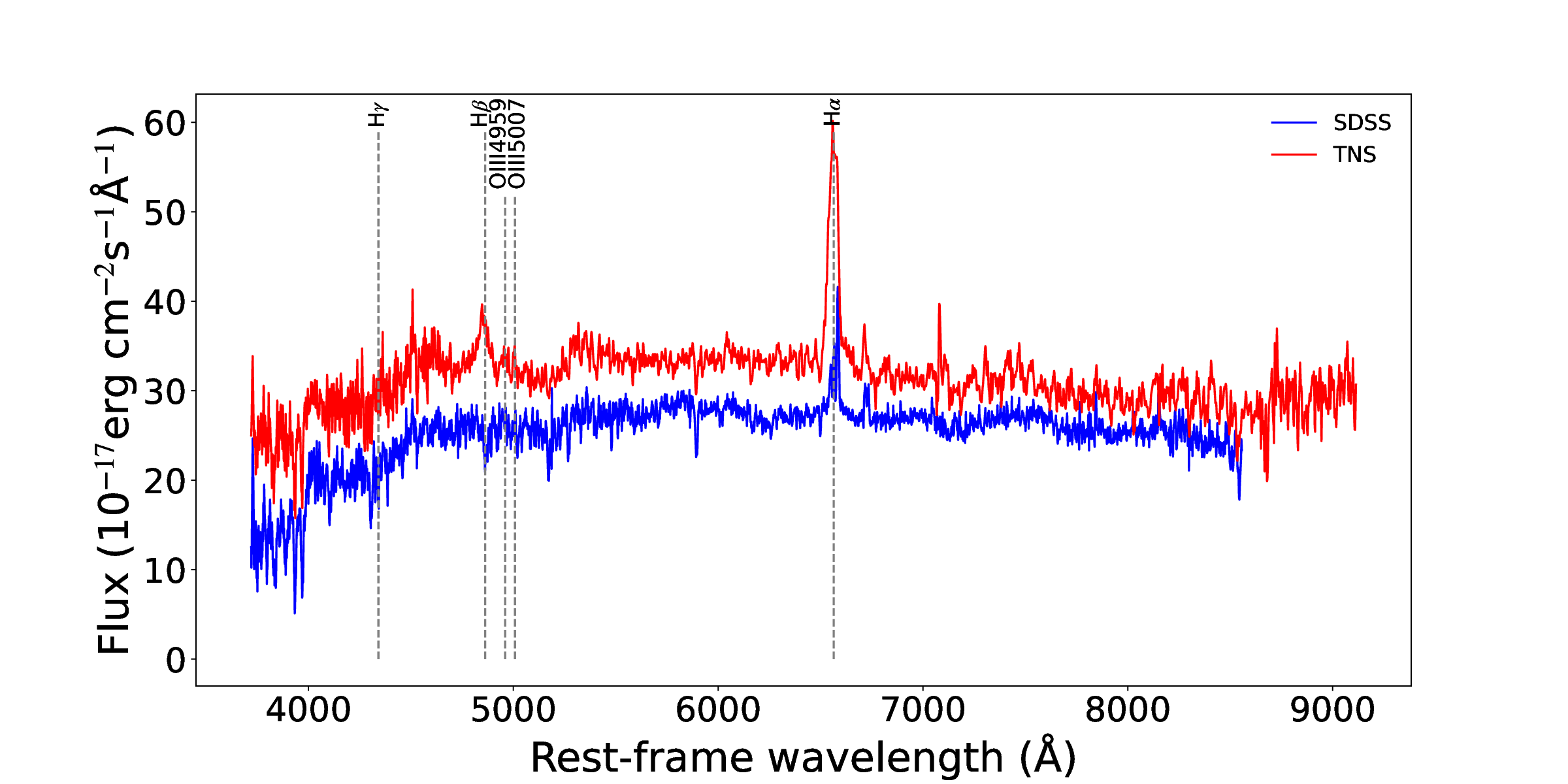}
			\caption{Same as Figure~\ref{fig:src1} for J1127+3546.}
			\label{fig:src2}
	\end{figure*}

\begin{figure}
\centering
\includegraphics[width=0.89\linewidth]{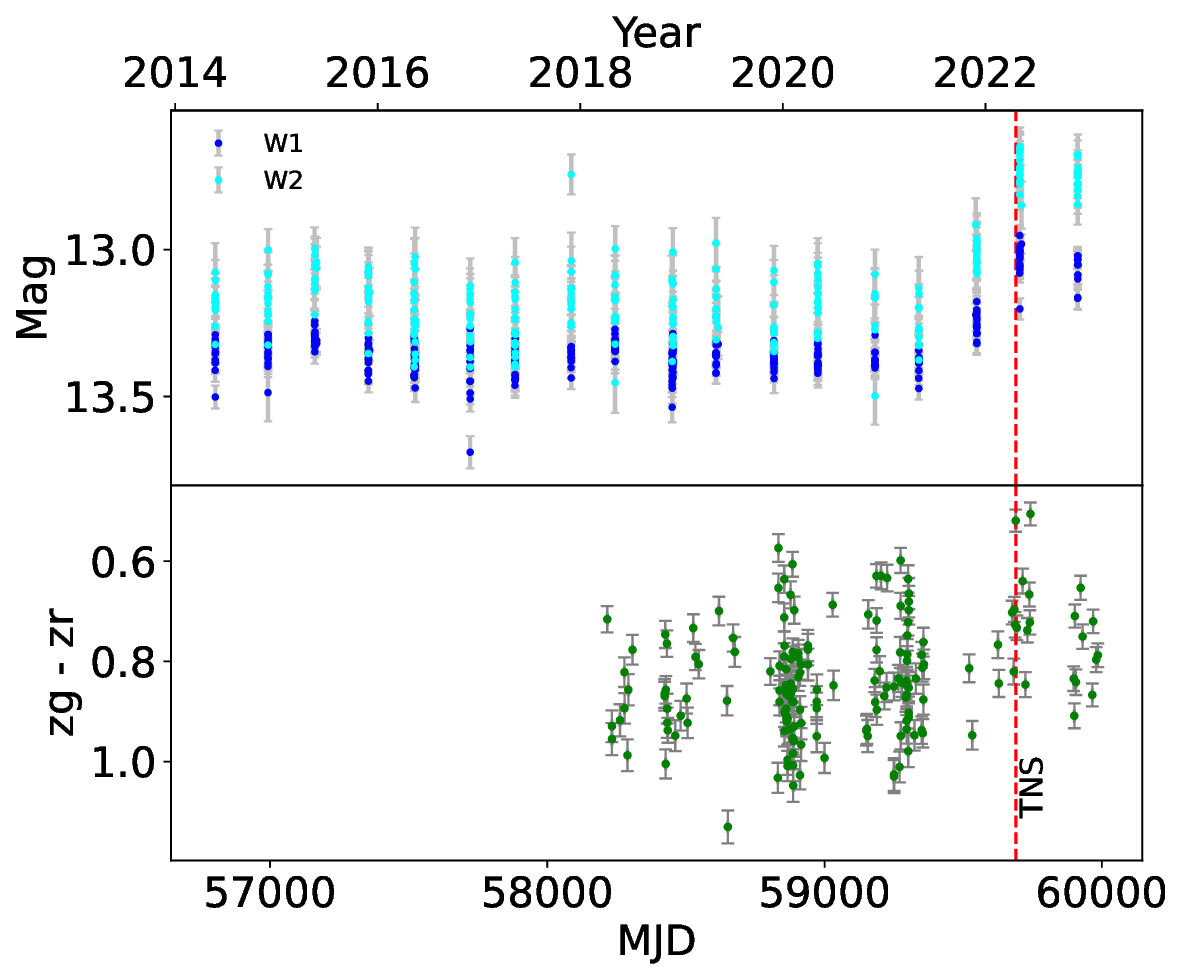}
	\caption{Same as Figure~\ref{fig:c1} for J1127+3546. The dashed line
	marks the TNS spectrum taking time.
		\label{fig:c2}}
\end{figure}
	
\subsection{J1127+3546}

This source also had a flat $V$-band light curve in the past, but
appears to have become highly variable in more recently optical light curves 
(Figure~\ref{fig:src2}). In particular, at the observation of the TNS spectrum, 
its $zg$ and $zr$ appeared to have risen by $\sim -$0.5\,mag, similarly
seen in the ATLAS $ac$ and $ao$ light curves.
The SDSS spectrum taken approximately 18 years ago shows the presence of a 
broad (but relatively faint) $H_\alpha$ component, while the absence of a 
similar H$_{\beta}$ component
(Table~\ref{tab:line}; see also Figure~\ref{fig:sf2}). This type of the 
spectrum suggests the possibility of the source being a Seyfert 1.8 
Galaxy, according to those characteristics outlined in \citet{csd+08}. 
Then based on the TNS spectrum reported by \citealt{pda+22}, the source was 
classified as a Seyfert 1 Galaxy, since both $>$2500\,km\,s$^{-1}$ FWHM
H$_\alpha$ and H$_{\beta}$ components were detected. In addition, the 
H$_{\alpha}$ component appeared much stronger than that in the SDSS spectrum. 
Using the TNS spectrum,
$M_{\rm BH}$ was estimated to be $\sim 10^{7.2}\,M_{\odot}$.

We also constructed a $zg-zr$ color diagram for the source, which is shown
in Figure~\ref{fig:c2}. As can be seen, the color values have a large spread 
in the range of 1.0--0.6. Thus, this source not only showed large
optical flux variations but also large color variations. 
Although there is a large spread in values, we tested to fit the colors 
as a function of $zr$, and obtained $zg - zr\propto 0.2zr$. 
However, this possible
correlation is weak, as the Spearman's rank correlation coefficient is only 
0.2 (corresponding to a $p$-value of 0.8\%). By comparison, the coefficient
for the correlation in J0113 (and J1223 below) is 0.61 (0.74).

It would be interesting to know what characteristics the spectrum of this
source had at the start of the ZTF light curves, where the large
flux and color variations occurred. Additionally, 
a small rise, with magnitude changes of 
$\sim -0.2$, can also be seen in the MIR bands at the time when
the TNS spectrum was taken.

\subsection{J1223+0645}

Overall, the variation features of this source, as seen in its multi-band 
light curves, appear similar to those of J0113. However, different from
J0113, the CRTS $V$-band light curve of this source also shows variations. 
Approximately from MJD~58200, 
the ZTF light curves have a $\sim -1$\,mag rise before reaching the maximum,
at which point our LJT spectrum was taken. The MIR fluxes accompanied
the optical rise. The SDSS spectrum for comparison was
taken nearly 18 years ago. While a weak, broad H$_{\alpha}$ component was 
required in the fitting of the SDSS spectrum 
(Table~\ref{tab:line}; Figure~\ref{fig:sf3}), our
LJT spectrum shows a similarly broad but significantly stronger H$_{\alpha}$ 
component.
The key difference that reveals the CL feature is the broad H$_{\beta}$ 
component, as no such a component was seen in the SDSS spectrum.
Based on the LJT spectrum,
$M_{\rm BH}$ was estimated to be $\sim 10^{8.8}\,M_{\odot}$.
	\begin{figure*}
			\includegraphics[width=0.69\linewidth]{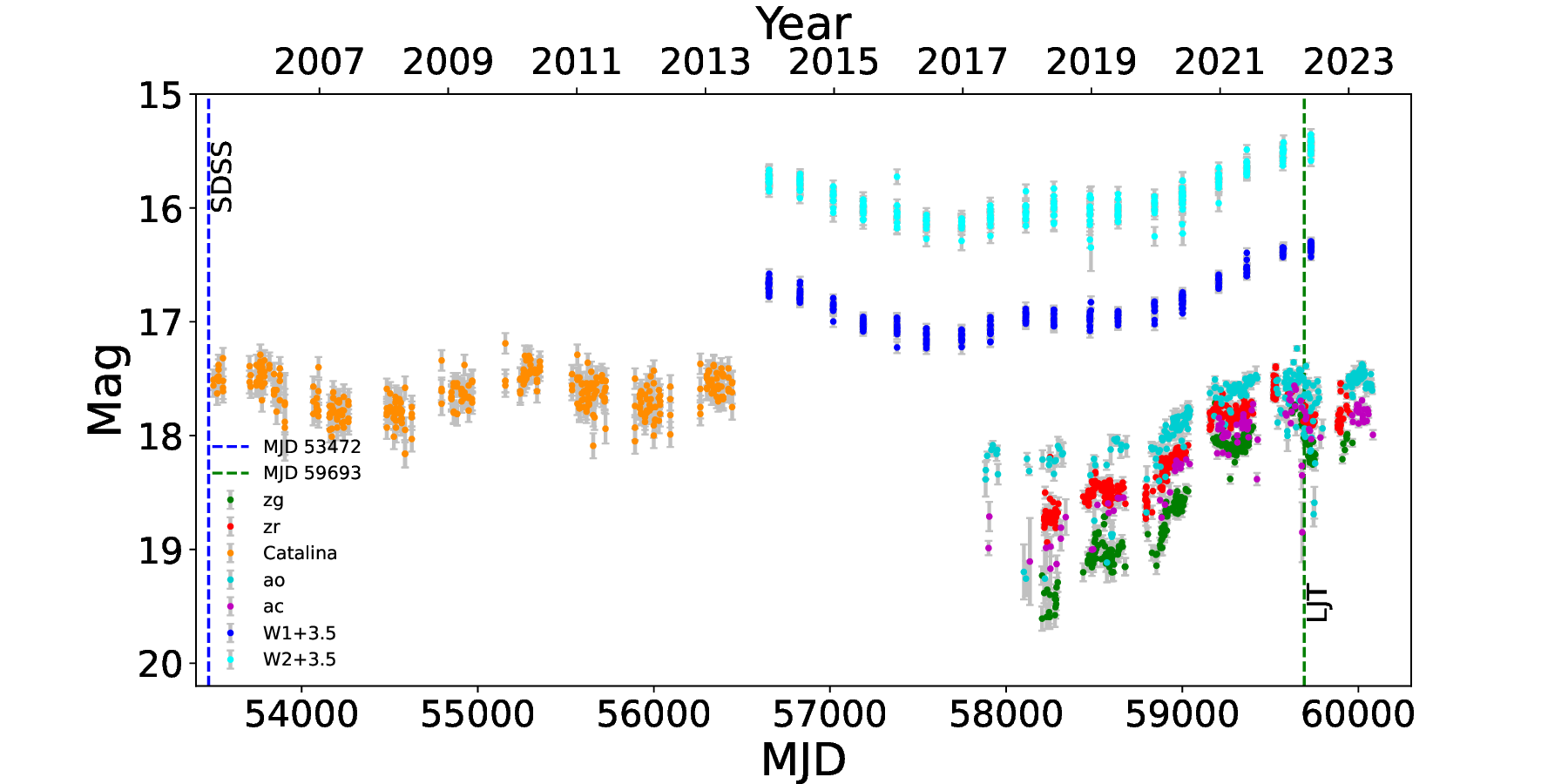}
			\includegraphics[width=0.69\linewidth]{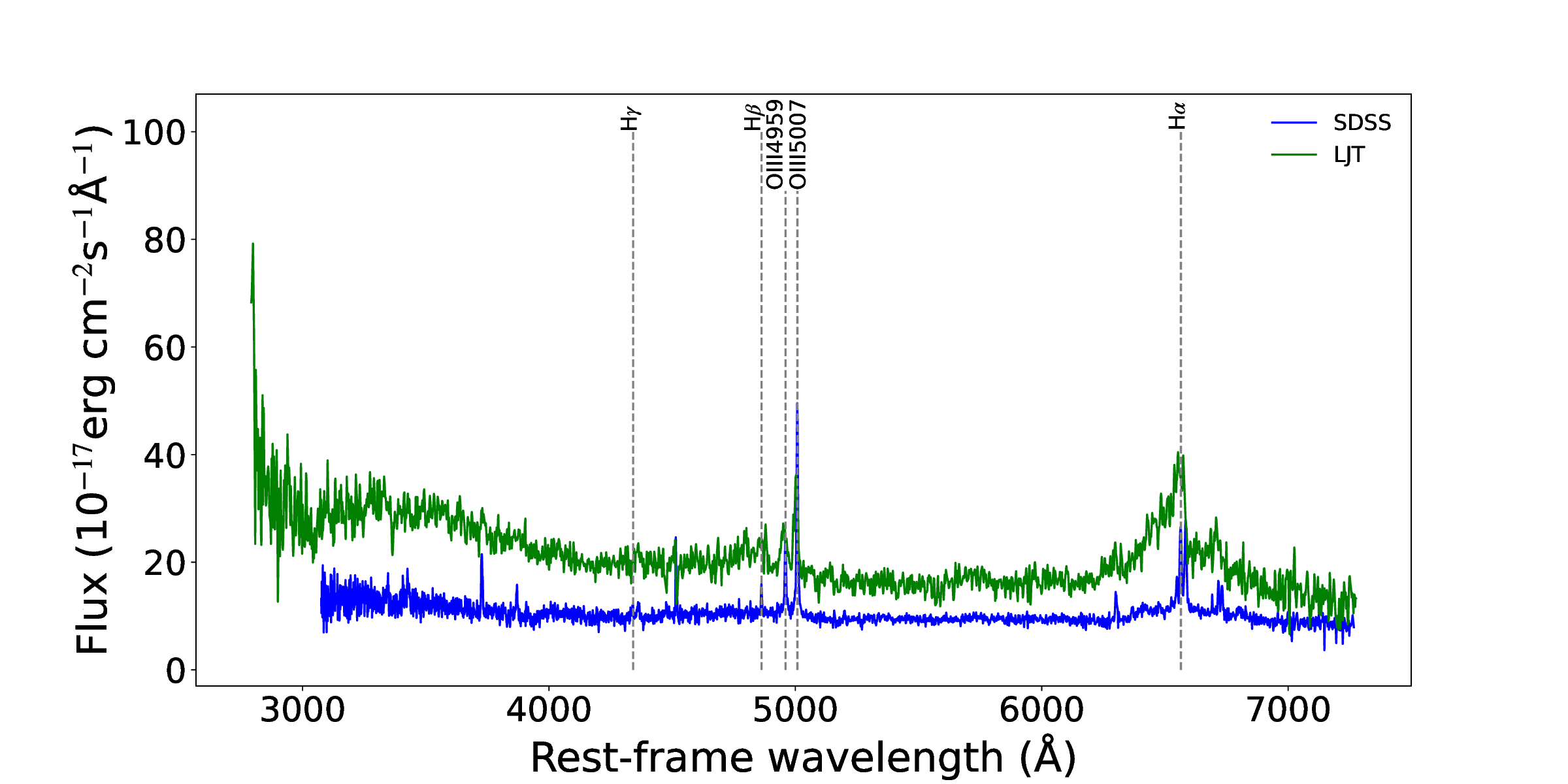}
			\caption{Same as Figure~\ref{fig:src1} for J1223+0645.}
		\label{fig:src3}
	\end{figure*}

\begin{figure}
\centering
\includegraphics[width=0.89\linewidth]{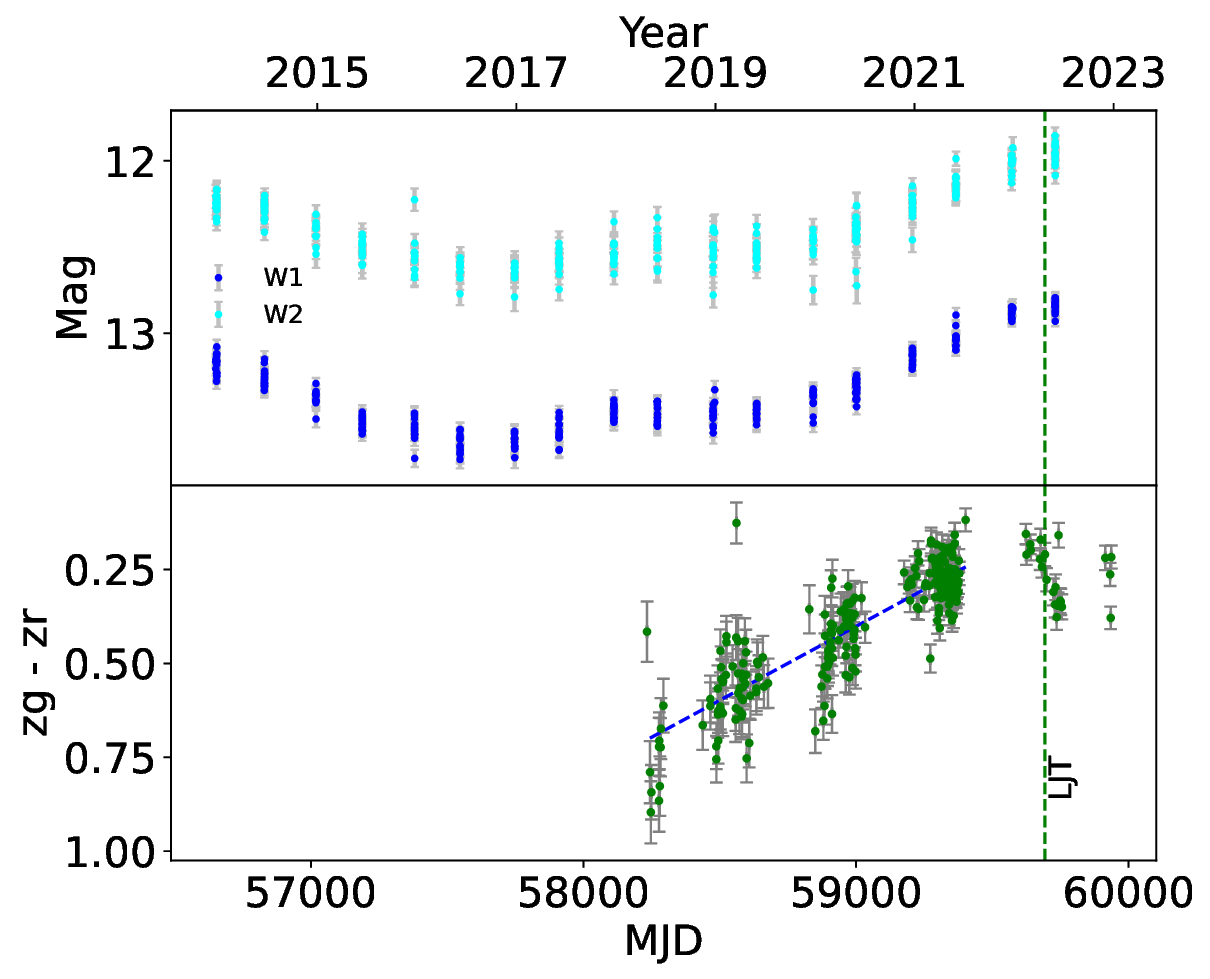}
	\caption{Same as Figure~\ref{fig:c1} for J1223+0645. The vertical
	dashed line
	marks the LJT spectrum taking time. In addition, a rate of 
	$-3.94\times 10^{-4}$\,mag\,day$^{-1}$ (marked by the blue
	dashed line in the {\it bottom} panel) can be derived for
	the data points approximately before MJD~59500.
		\label{fig:c3}}
\end{figure}

A $zg-zr$ color diagram was constructed to show the color changes, as the
$zg$ light curve had a faster rise than the $zr$ one (Figure~\ref{fig:c3}). 
For this source,
excluding two obvious outliers (which were likely real changes due to
large $zg$ magnitude changes, checked by us on the images and the data points),
the color values rise at a constant rate
of $-3.94\pm0.10\times 10^{-4}$\,mag\,day$^{-1}$, reaching a maximum 
before MJD~59500. After the maximum, our LJT spectrum was taken
(see Figure~\ref{fig:c3}).
A brighter-and-bluer pattern can also be well established, with
$zg - zr\propto 0.36 zr$ (see Figure~\ref{fig:cm}).
Similar to J0113, the MIR fluxes increased during the optical-rise time 
period, and the magnitude changes were $\sim -0.5$.

\begin{figure*}
	\includegraphics[width=0.69\linewidth]{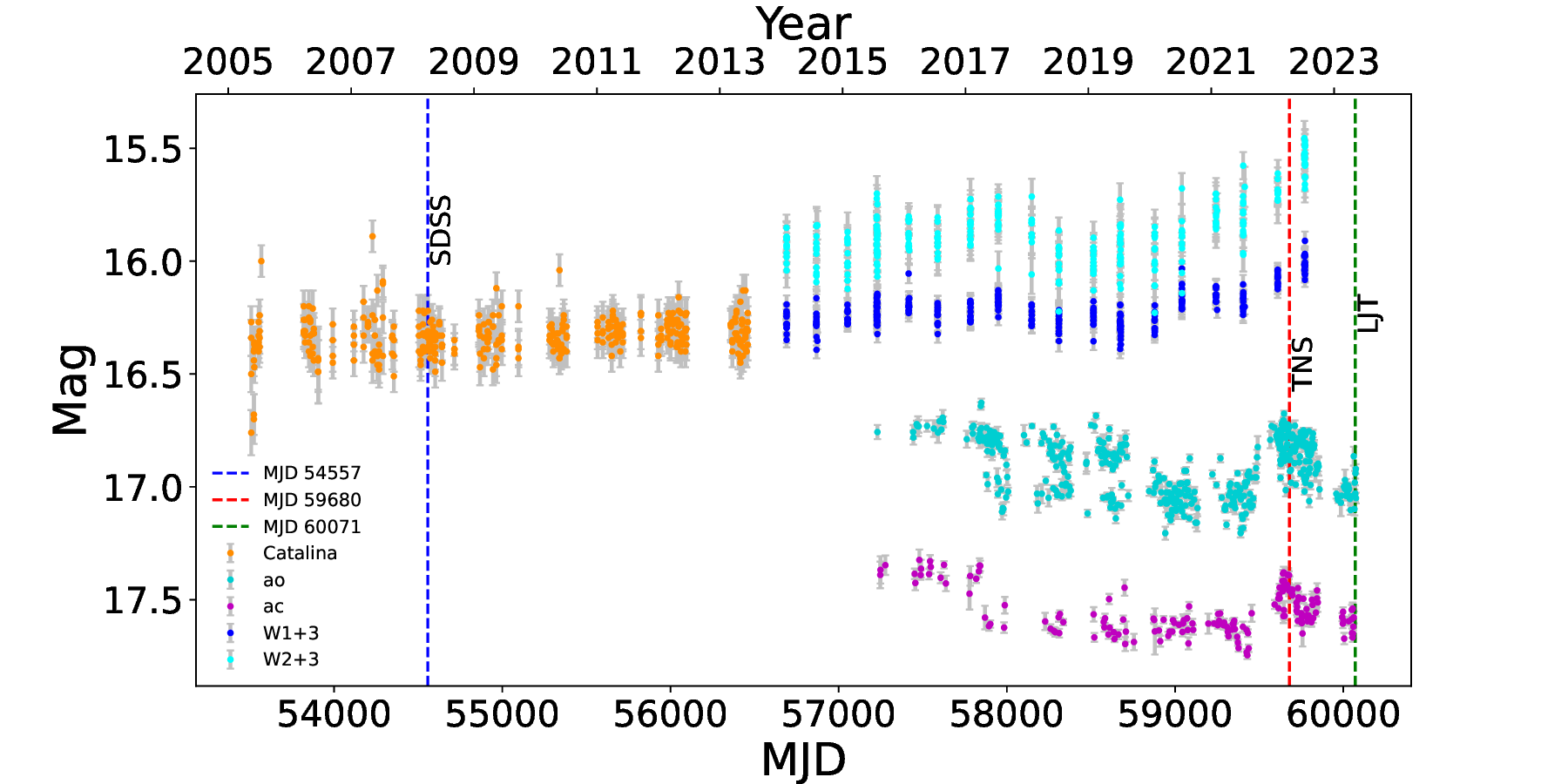}
	\includegraphics[width=0.69\linewidth]{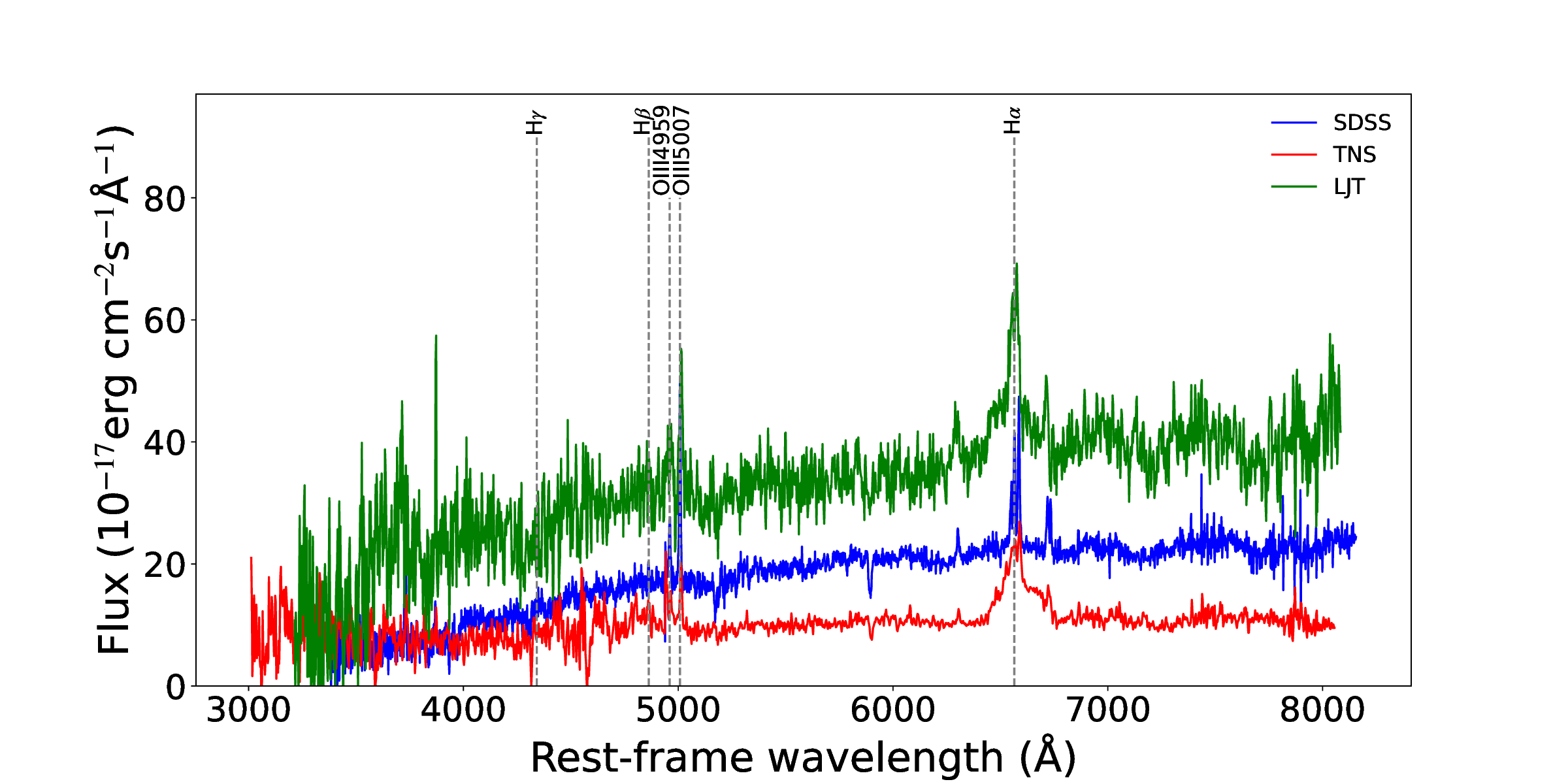}
	\caption{Same as Figure~\ref{fig:src1} for J1513+1759.}
	\label{fig:src4}
	\end{figure*}

\subsection{J1513+1759}

This source's $V$-band light curve appears flat and with little variation,
but after MJD~57000, both the optical and MIR light curves
show significant variations.
Because there were only a few data points from ZTF, only ATLAS $ac$ and
$ao$ light curves are shown in Figure~\ref{fig:src4}.
The TNS
spectrum, taken on MJD~59680, shows the detection of a broad H$_{\alpha}$ 
component as well as the weak detection of a broad H$_{\beta}$ component
(Table~\ref{tab:line}; Figure~\ref{fig:sf4}). In contrast to
the SDSS spectrum taken $\sim$15\,yr ago that does not show any broad 
components, the newly detected features indicate the CL behavior of 
the source.  It should be noted that at the time the TNS 
spectrum was taken, the optical light curves were at a local variation peak. 
Our LJT spectrum was taken $\sim$1\,yr later, when the optical light curves
reverted to a lower brightness level, but it still contains the broad
H$_{\alpha}$ component. A broad H$_{\beta}$ was not detected, but 
the absence of the component could be due to the low signal-to-noise detection 
of the source spectrum. 

We did not construct a color diagram for the optical variations of this source,
since the ATLAS bands are much wider than typically used ones \citep{tdh+18}.
	
	\begin{table*}
			\centering
			\caption{Measurements from fitting the spectra with PyQSOFit}
			\label{tab:line}
			\begin{tabular}{lccccc}
				\hline
Line & (J0113) SDSS  & TNS  & LJT  & (J1127) SDSS   & TNS  \\ 
			\hline
	\multicolumn{5}{l}{H$\alpha$\quad broad}\\
		\hline 
FWHM  & 0 & 5885$\pm$106 & 5956$\pm$157 & 5146$\pm$384 & 2889.4$\pm$9.2  \\
EW   & 0 & 153.5$\pm$1.6    & 165.6$\pm$3.8  & 233$\pm$19  & 447$\pm$10 \\
Flux & 0 & 3900$\pm$41  & 1647$\pm$38& 337$\pm$26  & 1038$\pm$24 \\
Peak   & 0 & 6549.5$\pm$1.1    & 6569.2$\pm$1.5 & 6552$\pm$5.0  & 6556.98$\pm$0.49  \\
				\hline
\multicolumn{5}{l}{H$\alpha$\quad narrow}\\
	\hline
FWHM & 443.0$\pm$6.2   & 551$\pm$21  & 709$\pm$35 & 394$\pm$17 & 826$\pm$14 \\
EW   & 20.84$\pm$0.49  & 13.0$\pm$0.68    & 17.4$\pm$1.3  & 33.7$\pm$4.4 & 113.9$\pm$5.0 \\
Flux & 162.1$\pm$3.8 & 331$\pm$17 & 172$\pm$13 & 49.3$\pm$6.5 & 266$\pm$12 \\
				\hline
\multicolumn{5}{l}{H$\beta$\quad broad}\\
	\hline
FWHM  & 0 & 5642.5$\pm$2.1 & 4656$\pm$293 & 0 & 2526$\pm$61  \\
EW   & 0 & 37.4$\pm$0.9    & 20.2$\pm$1.9  & 0 & 36.98$\pm$0.85  \\
Flux & 0 & 1169$\pm$28 & 240$\pm$22 & 0 & 297.8$\pm$6.8 \\
	\hline
\multicolumn{5}{l}{H$\beta$\quad narrow}\\
	\hline
FWHM & 495.9$\pm$9.7 & 578$\pm$59 & 991$\pm$57 & --- & 319$\pm$50 \\
EW   & 3.91$\pm$0.48  & 1.83$\pm$0.53    & 1.83$\pm$0.91   & --- & 1.90$\pm$0.69 \\
Flux & 30.4$\pm$3.7 & 57$\pm$17  & 22$\pm$11 & ---  & 15.1$\pm$5.5 \\
	\hline
MJD  & 57282          & 59410            & 59872  & 53468  & 59690  \\
				\hline
				log($M_{BH}$/$M_{\odot}$)& ---  & 8.63$^{+0.06}_{-0.07}$  & 8.23$^{+0.12}_{-0.13}$  & --- & 7.18$\pm$0.03 \\
				\hline        
			\end{tabular}\\
		    \begin{tabular}{lccccc}
		    	\hline
  Line   & (J1223) SDSS & LJT & (J1513) SDSS   & TNS  & LJT   \\ 
		    	\hline
\multicolumn{5}{l}{H$\alpha$\quad broad}\\
		    	\hline 
    FWHM    & 11973$\pm$3333 & 14125$\pm$2350 & 0 & 8872$\pm$571 & 8189$\pm$62 \\
  EW  & 82.9$\pm$8.1 & 274.4$\pm$5.3 & 0 & 1141$\pm$43  & 413.0$\pm$3.0 \\
 Flux & 970$\pm$94 & 6237$\pm$120 & 0 & 1728$\pm$64  & 2105$\pm$15  \\
  Peak & 6548$\pm$19 & 6507$\pm$17 & 0 & 6551.1$\pm$2.2   & 6547.93$\pm$0.83  \\
		    	\hline
\multicolumn{5}{l}{H$\alpha$\quad narrow}\\
	    	\hline
 FWHM & 392$\pm$14 & 855$\pm$168 & 264$\pm$10 & 604$\pm$75 & 722.6$\pm$7.4  \\
 EW  & 15.03$\pm$0.97 & 15.4$\pm$1.0 & 147.9$\pm$4.7  & 46.6$\pm$6.9  & 54.12$\pm$0.79 \\
 Flux & 176$\pm$11 & 351$\pm$23 & 97.0$\pm$3.1  & 71$\pm$11  & 276.0$\pm$4.0  \\
		    	\hline
		    	\multicolumn{5}{l}{H$\beta$\quad broad}\\
		    	\hline
   FWHM  & 0 & 7311$\pm$1032 & 0 & 5599.6$\pm$5.1 & 0\\
   EW & 0 & 50.02$\pm$0.81 & 0 & 87.2$\pm$3.2  & 0 \\
    Flux & 0 & 1262$\pm$20 & 0 & 390$\pm$15 & 0 \\
		    	\hline
		    	\multicolumn{5}{l}{H$\beta$\quad narrow}\\
		    	\hline
FWHM & 392$\pm$13 & 1206$\pm$419 & 473.1$\pm$7.5 & 818$\pm$125 & 972$\pm$15 \\
EW & 4.02$\pm$0.18  & 6.6$\pm$2.6 & 23.1$\pm$4.8  & 9.4$\pm$3.1  & 91.8$\pm$2.7 \\
Flux & 1.0$\pm$2.3   & 130.0$\pm$4.0 & 23.3$\pm$4.8   & 41$\pm$13   & 116.8$\pm$3.4 \\
		    	\hline
   MJD & 53472 & 59693 & 54557 & 59680 & 60071 \\
		    	\hline
			    log($M_{BH}$/$M_{\odot}$)& --- & 8.75$^{+0.16}_{-0.18}$ & --- & 8.45$^{+0.19}_{-0.23}$ & --- \\
		    	\hline        
		    \end{tabular}
	    \begin{tablenotes}[]
	    \item Notes. 1) Full Width at Half Maximum (FWHM), Equivalent Width (EW), flux, and peak measurements are in units of km\,s$^{-1}$, angstrom (\AA), $10^{-17}$ erg\,cm$^{-2}$\,s$^{-1}$, and angstrom (\AA), respectively. 2) Approximate systematic uncertainties of 6\%, 3\%, and 9\% for the TNS spectra of J0113, J1127, and J1513, respectively, and 10\%, 9\%, and 17\% for the LJT spectra of J0113, J1223, and J1513, respectively, should be considered for the measurements of the line features. 3) The systematic uncertainties are included to
		   estimate the black hole masses.
	    \end{tablenotes}            
		\end{table*}
	   
\section{Discussion}
\label{sec:dis}

Selecting significantly variable AGN from the ZTF as well as the ATLAS survey 
data, we have found four sources with the CL feature by comparing the spectra 
from the archival data and our own targeted spectroscopic observations. 
In the respective time periods wherein significant flux variations occurred,
the spectra of the sources show the appearance of a broad H$_\alpha$ 
(in the cases of J0113 and J1513) or a stronger one (in the cases of J1127 
and J1223).
Most of the spectra also show the appearance of a broad, but relatively weak, 
H$_\beta$ component.

For J0113, \citet{lmb+22} have identified it as a CLAGN, and the TNS and LJT
spectra in this work have confirmed their identification. Their spectrum 
was taken at $\sim$MJD~59520, in between those of the TNS and LJT
and around the peak of the optical rise. The broad H$_\beta$ component 
in their spectrum 
appeared to be stronger than that in the TNS or LJT spectrum, which is probably
not a coincidence and likely has physical connections. For J1513, although 
the TNS and LJT spectra are not of
good quality, the absence of a broad H$_\beta$ in the SDSS spectrum, as well as
its possible appearance and re-absence in the TNS and LJT spectra, show a
similarity to those CL cases such as Mrk~590 \citep{ddc+14}, 
Mrk~1018 \citep{mhc+16},
and in particular NGC~4156 \citep{tlc22}. In addition to the appearance
of a broad H$_\alpha$ component in 2019, NGC~4156 was found to have a
relatively weak, broad H$_\beta$ component, one that weakened further 
in 2022. The possible weakening of the broad H$_\beta$ component in J1513 
occurred over a $\sim$400\,day period, which is marked by the TNS and LJT
spectra.

Another notable feature found in the four AGN is the MIR brightening that 
accompanied the respective optical flux increase (or variations in J1513). 
It has been found that this
could be a common feature of CL activity \citep{she+17} and may be
used as an indicator for finding CLAGN \citep{swj+20}.
    
The origin of the CL phenomena remains enigmatic. However, the correlated
MIR variations can often help exclude the variable obscuration,
one broad category of the scenarios often discussed for the CL phenomena,
as the cause \citep{she+17,rfg+18,ste+18}. 
On the other hand, detailed studies of
several CL cases have been able to limit potential causes to some
physical processes or changes, such as the state transition (which has been
well studied in black-hole X-ray binaries; \citealt{nd18,ady+20}) or
disk instabilities \citep{rfg+18,ste+18,scb+20} in the inner part of 
the accretion gas, which is another broad category of the possible scenarios. 
Because of the
comparatively short timescales (months to years) of CL transitions,
the process-occurring regions are thus likely small and close to 
the SMBHs (e.g., \citealt{ste+18,scb+20}). 

Related to the second category of the possibilities, it is conceivable to
observe the flux variations that accompany CL transitions, maybe especially
in cases demonstrating a change from type~2 to type~1, as the physical 
processes that could induce the transitions would naturally produce
significant optical flux variations. Taking our four cases as an example,
they were previously classified as type~2 based on the SDSS spectra, but their 
relatively large flux variations attracted our attention to carry out 
follow-up studies. Moreover, for two of them, we have found 
a brighter-and-bluer feature, one that is commonly seen in type~1 AGN 
(see \citealt{rua+14} and references therein). Thus, we suspect that this 
brightness-color feature could be used as a criterion for finding CL sources. 
\begin{figure}
	\includegraphics[width=0.98\linewidth]{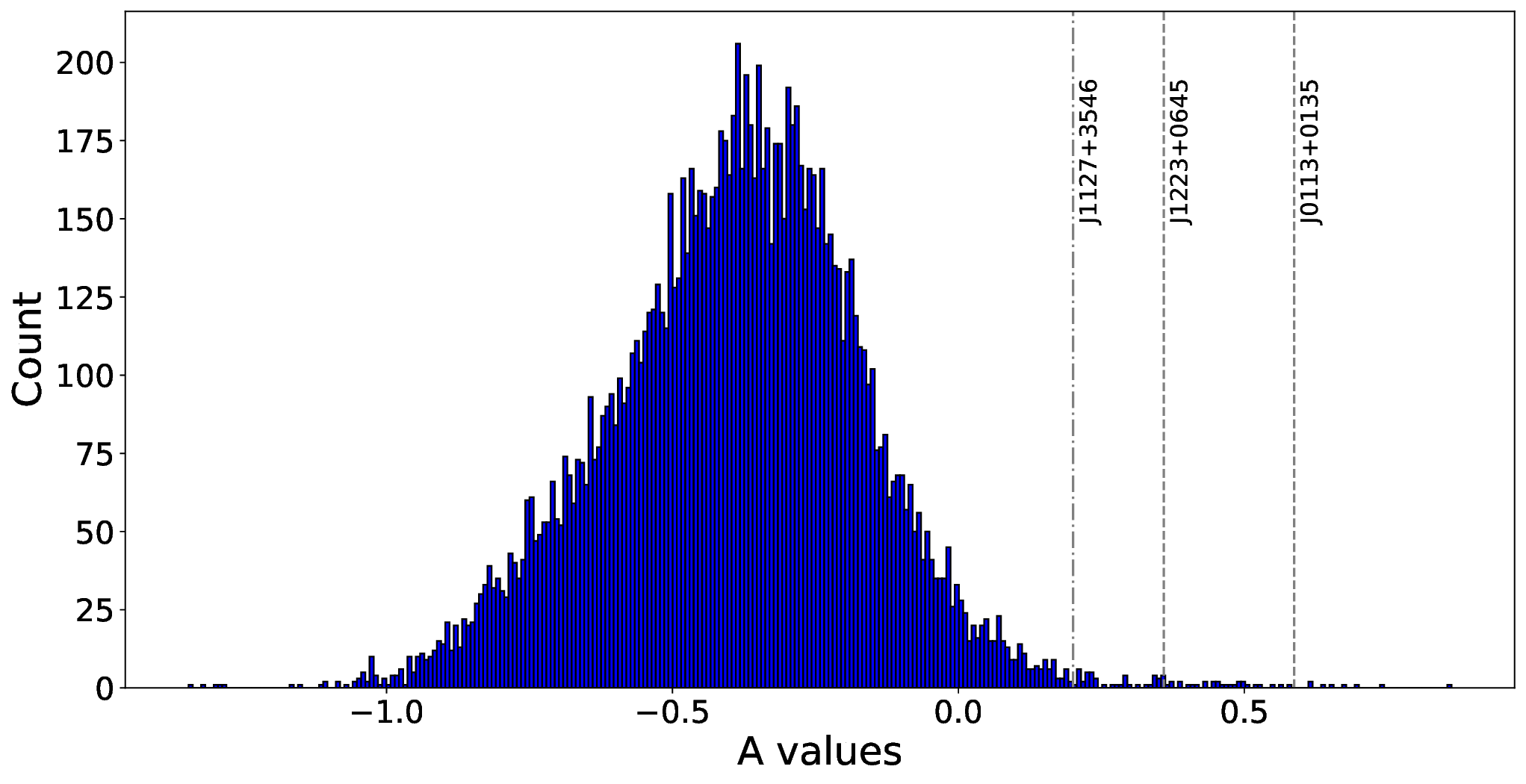}
	\caption{$k$-value distribution for 12614 SDSS type 2 AGN. 
	The values for two of the four sources in this work are marked
	by vertical dashed lines. The value for the weak correlation
	in J1127 is also marked, with a dash-dotted line.}
	\label{fig:kdist}
	\end{figure}

To roughly test this possibility,
we collected a large number of type 2 AGN among the SDSS sources and 
checked their
magnitude-color behavior, although it should be noted that the classifications
might not be reliable and could contain many complicated cases 
(see, e.g., \citealt{lop+23a}). We found 12614 SDSS type 2 AGN 
with ZTF data points greater than 43 in each of the $zg$ and $zr$ bands, where
we required the median signal-to-noise ratio {\tt snmedian} $\geq 10$ and 
redshift $z \leq$ 0.5 when querying the SDSS database (details will be 
reported elsewhere in L. Zhu et al., in preparation).
Sources with too few ZTF data points were excluded because 
sparse data points do not allow us to build a clear magnitude-color 
relationship.  We then fit $zg-zr$ colors (obtained from
the measurements within one day) as a function of $zr$,
$zg-zr\propto k\times zr$, where $k$ is the slope. The distribution
of the obtained $k$ values is shown in Figure~\ref{fig:kdist}. As can be seen, 
different from two of our cases,
whose $k$ values are also indicated in Figure~\ref{fig:kdist},
most of the sources in the sample we collected have $k$ values smaller than 0,
which implies that they probably had a brighter-and-redder behavior in 
the ZTF data.
In addition, we also checked sources with $k\geq 0.1$. There are 177 of them, 
among which 10 have already been reported to display the CL phenomena.
Thus, we suggest that this color-changing
behavior can likely be used for finding BEL-on CLAGN among the recorded type~2
AGN. For example, sources with $k>0$ in Figure~\ref{fig:kdist} could be 
candidate CL 
sources. A spectroscopy survey of them will reveal if the 
color-changing behavior could be a key in efficiently finding CLAGN among
these previously SDSS spectroscopically classified type~2 AGN.
    
\section*{Acknowledgements}

We thank referee for comments that help improve our work.
This work was
based on observations obtained with the Samuel Oschin Telescope 48-inch and 
the 60-inch Telescope at the Palomar Observatory as part of the Zwicky 
Transient Facility project. ZTF is supported by the National Science 
Foundation under Grant No. AST-2034437 and a collaboration including Caltech, 
IPAC, the Weizmann Institute for Science, the Oskar Klein Center at
Stockholm University, the University of Maryland, Deutsches 
Elektronen-Synchrotron and Humboldt University, the TANGO Consortium of 
Taiwan, the University of Wisconsin at Milwaukee, Trinity College Dublin, 
Lawrence Livermore National Laboratories, and IN2P3, France. Operations are 
conducted by COO, IPAC, and UW.

This work made use of data products from the Wide-field Infrared Survey 
Explorer, which is a joint project of the University of California, Los 
Angeles, and the Jet Propulsion Laboratory/California Institute of Technology, 
funded by the National Aeronautics and Space Administration.

This research is supported by the Basic Research Program of Yunnan Province 
No. 202201AS070005, the National Natural Science Foundation of China
(12273033), and the Original
Innovation Program of the Chinese Academy of Sciences
(E085021002).

\section*{Data Availability}
The data underlying this article will be shared on reasonable request to
the corresponding author.

\bibliographystyle{mnras}
\bibliography{cl}

\appendix
\section{Spectrum fitting with PyQSOFit}
\label{sec:app}
	\begin{figure*}
			\centering
			\includegraphics[width=0.86\linewidth]{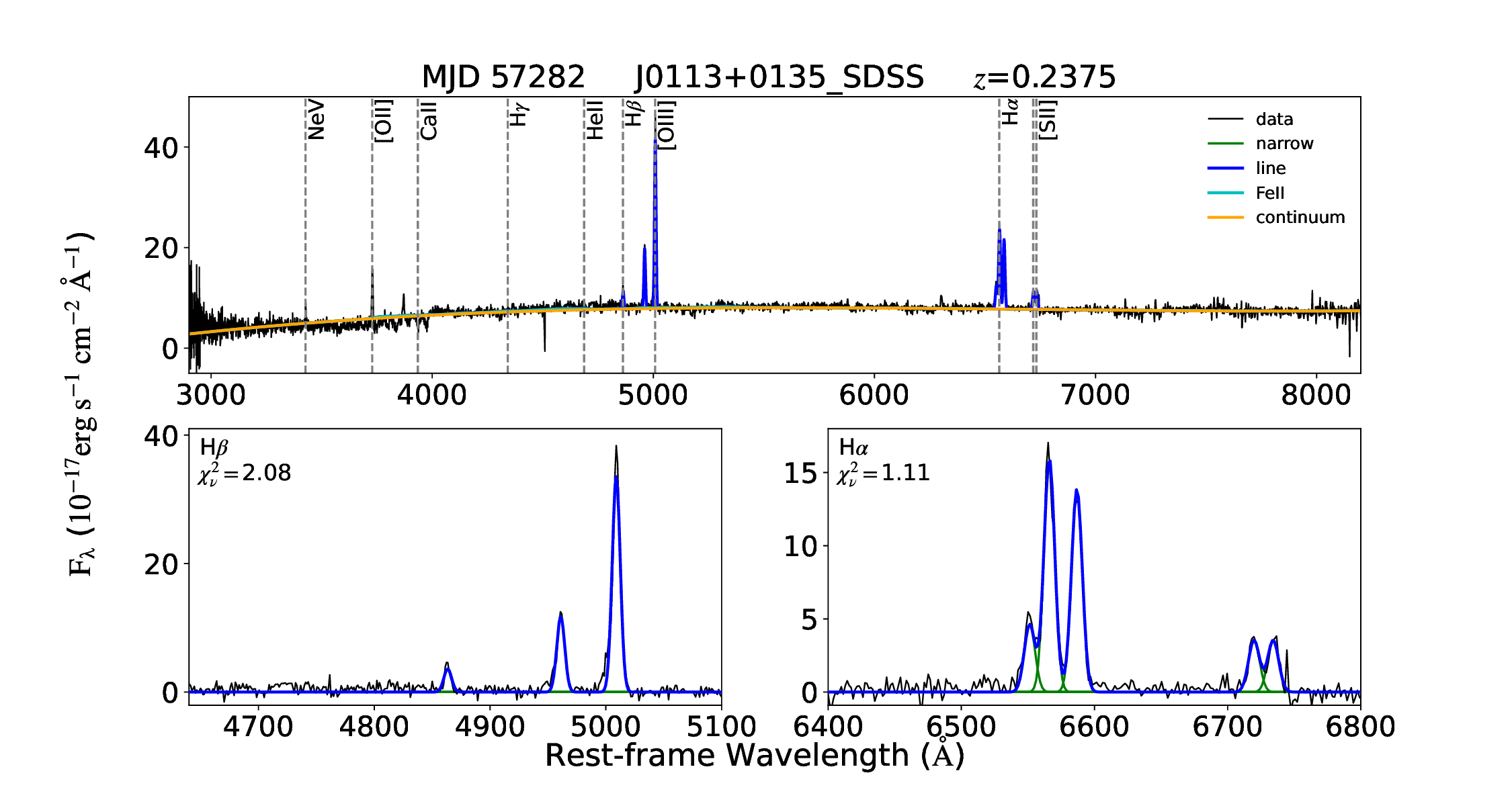}
			\includegraphics[width=0.86\linewidth]{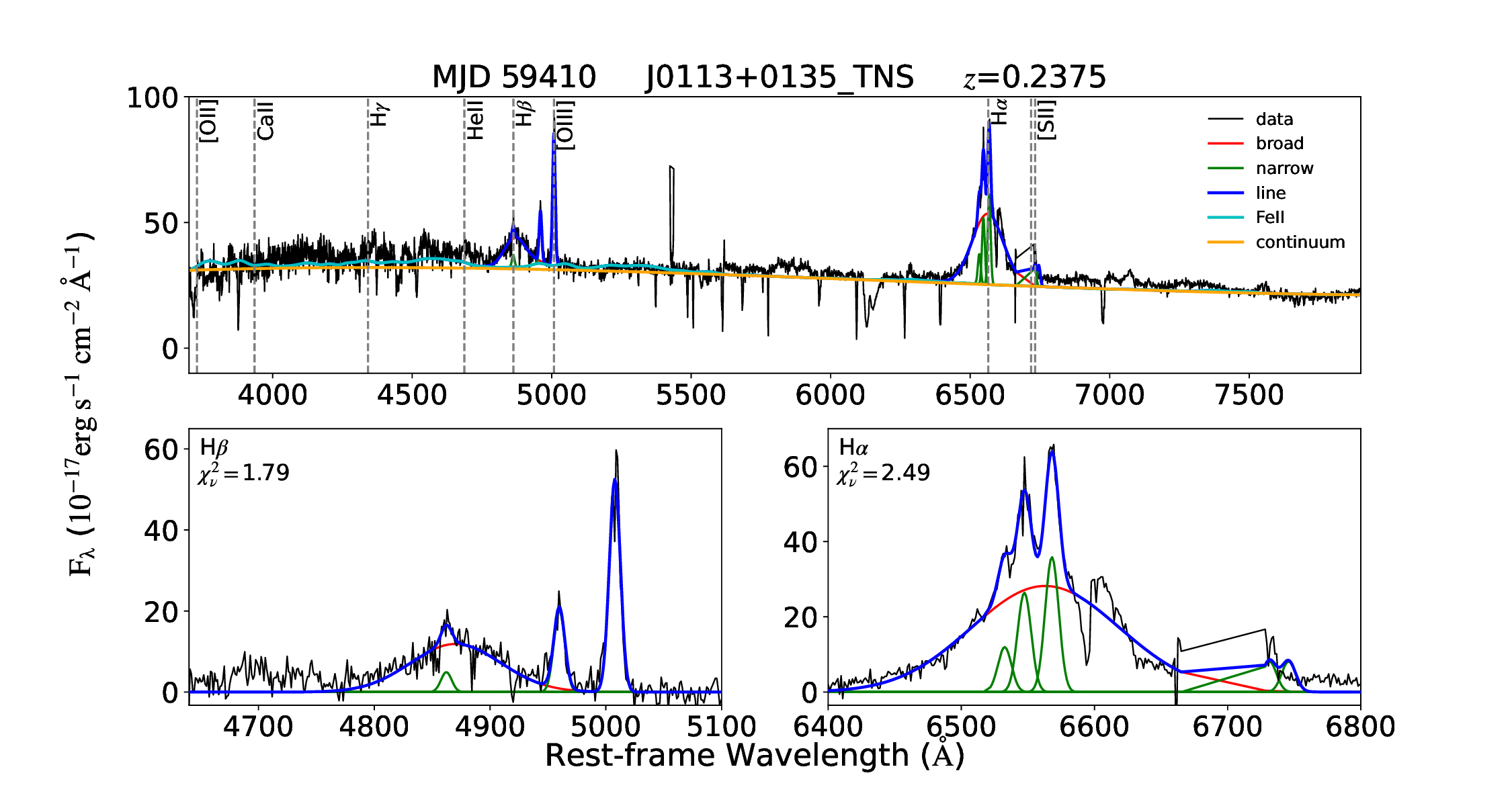}
			\includegraphics[width=0.86\linewidth]{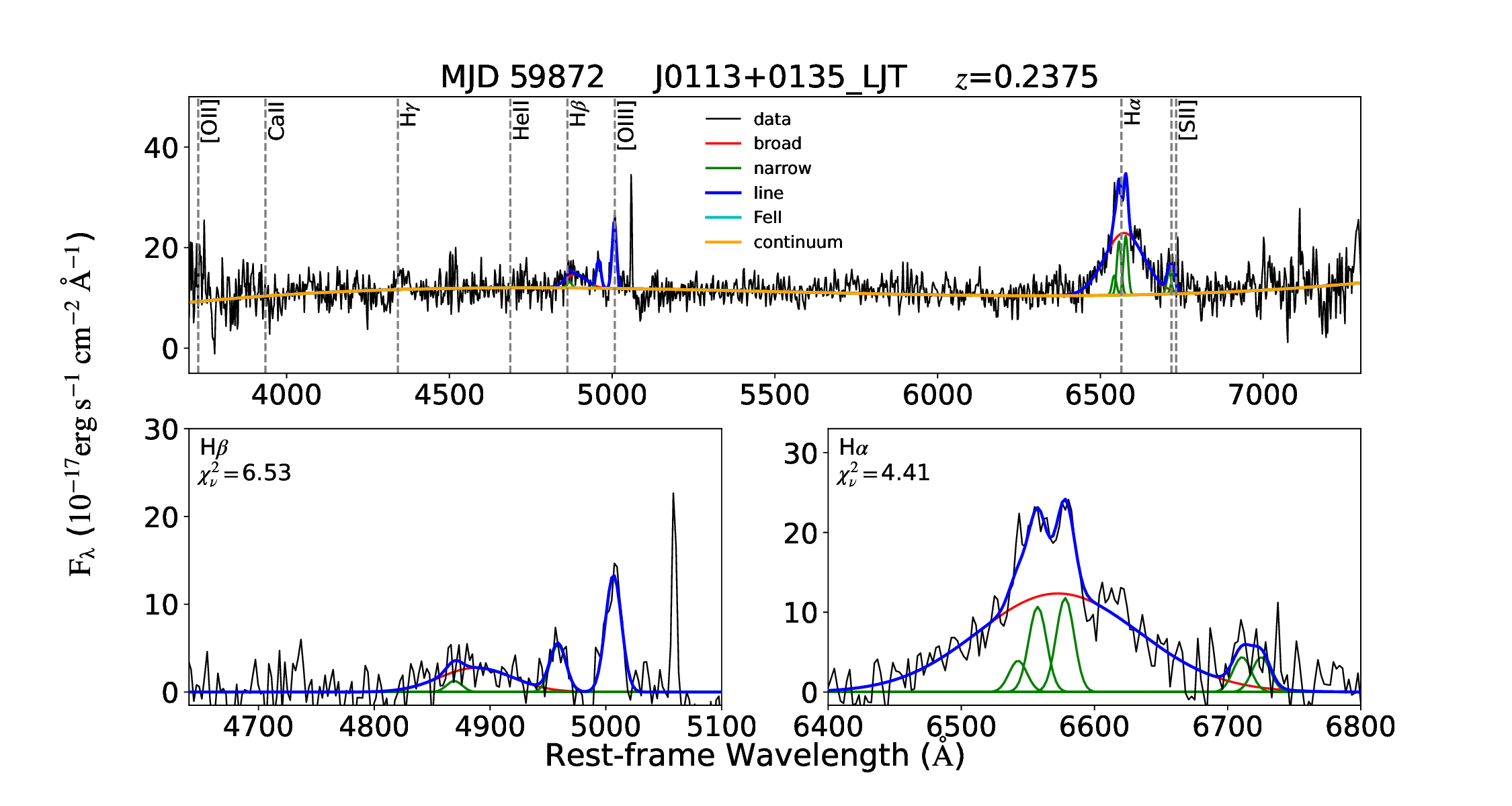}
	\caption{Spectral fitting for SDSS, TNS, and LJT spectra of J0113+0135.}
	\label{fig:sf1}
	\end{figure*}

	\begin{figure*}
			\includegraphics[width=0.86\linewidth]{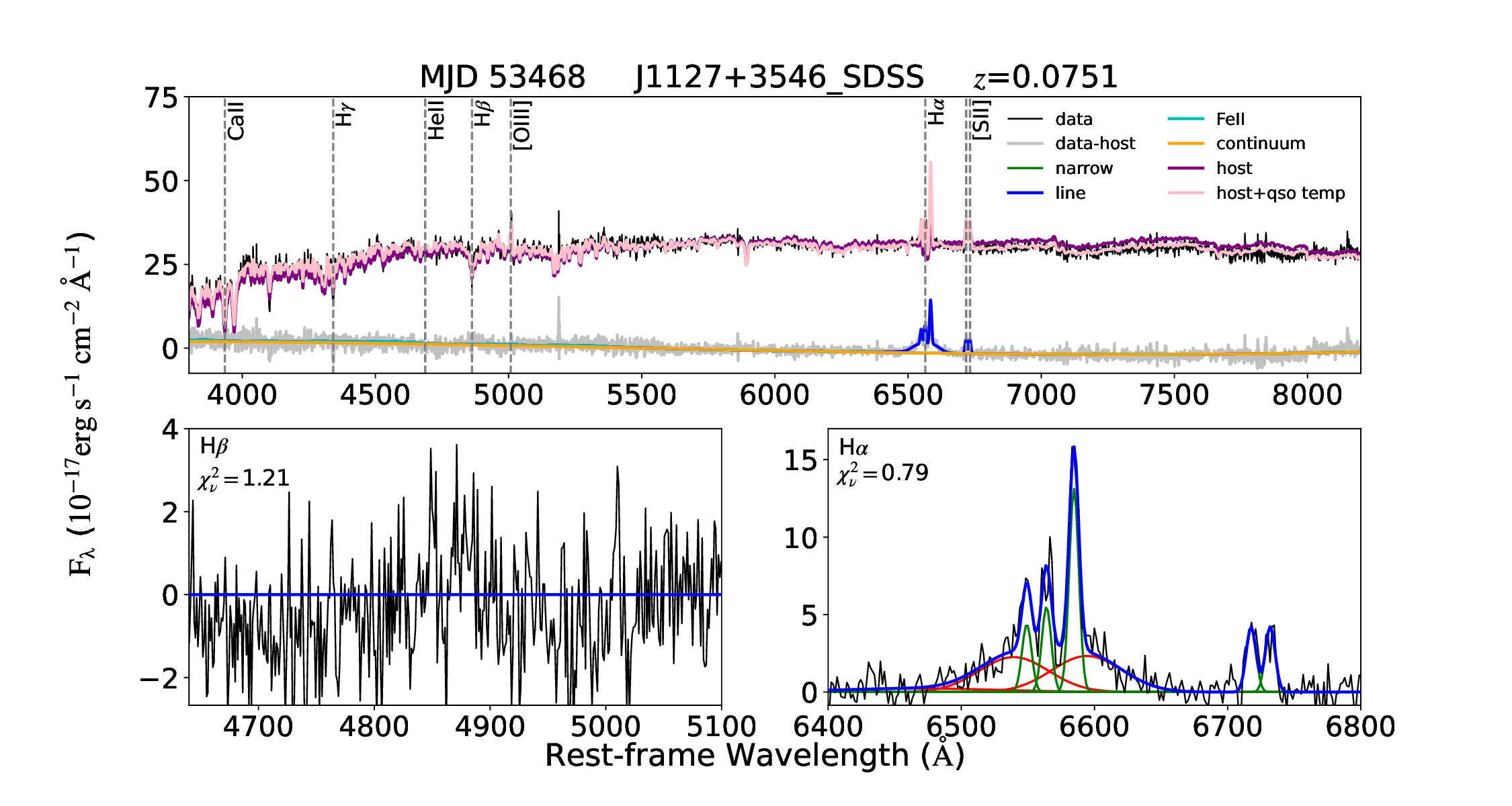}
			\includegraphics[width=0.86\linewidth]{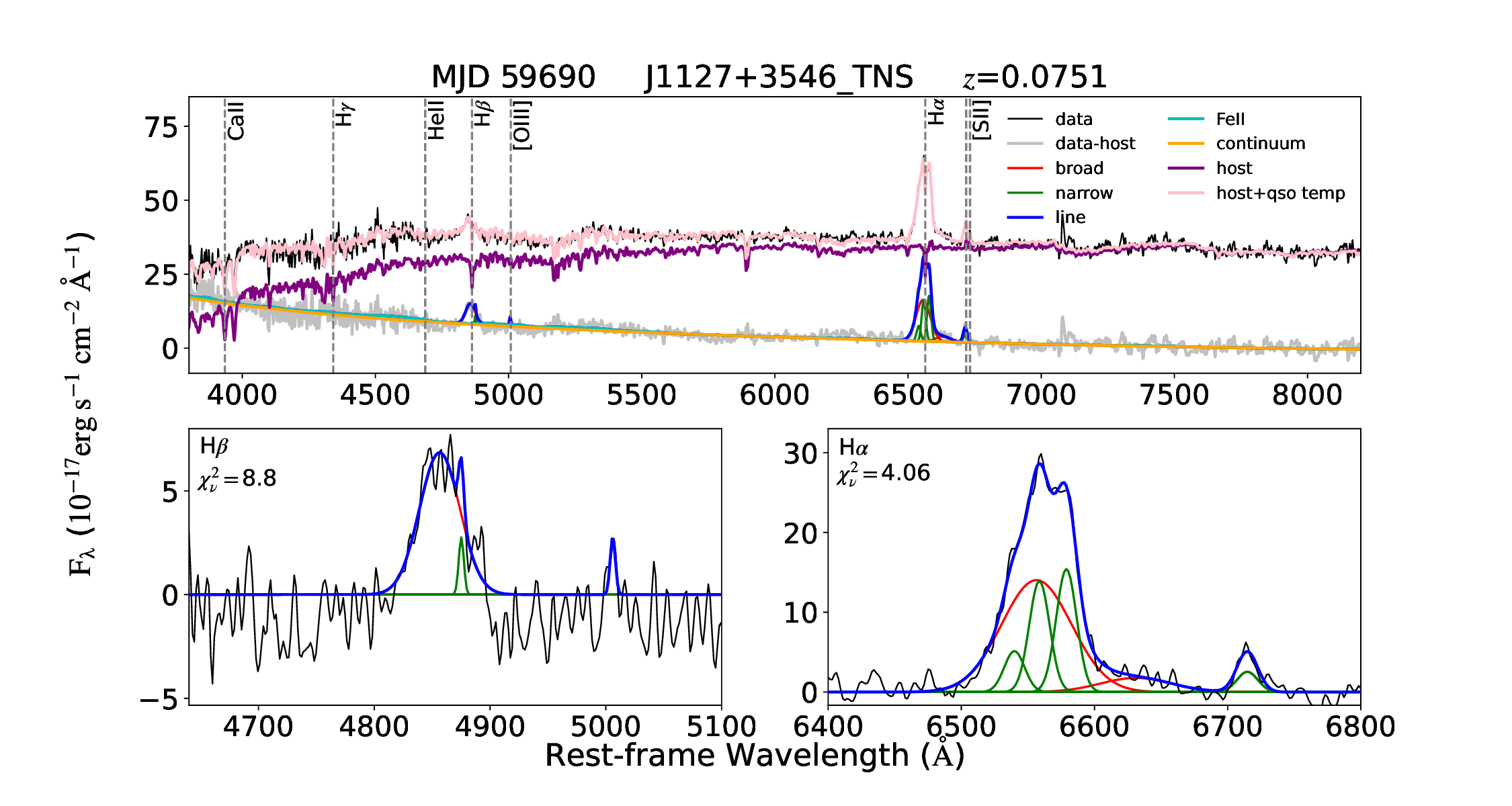}
	\caption{Spectral fitting for SDSS and TNS spectra of J1127+3546.}
	\label{fig:sf2}
	\end{figure*}

	\begin{figure*}
			\includegraphics[width=0.86\linewidth]{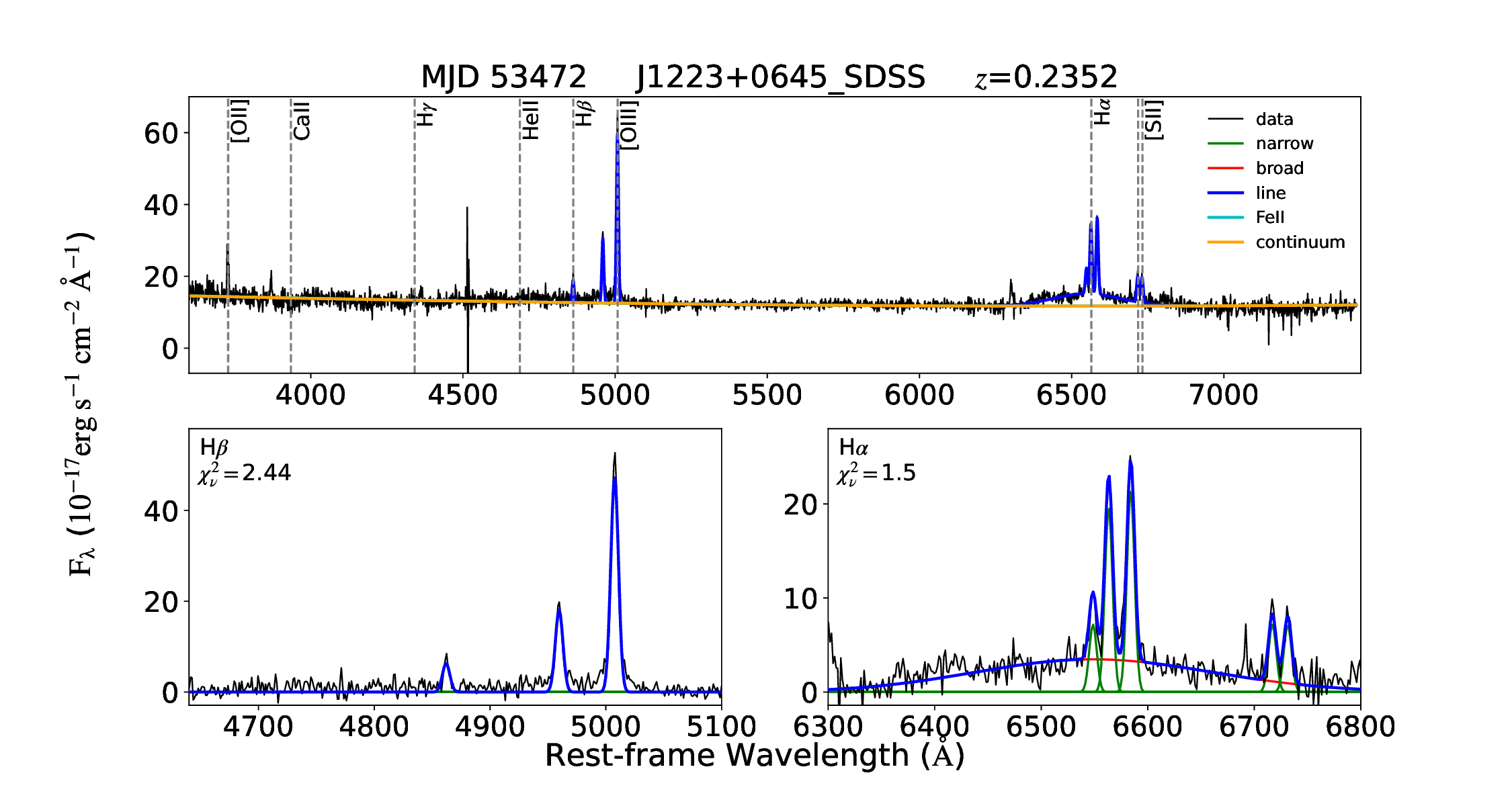}
			\includegraphics[width=0.86\linewidth]{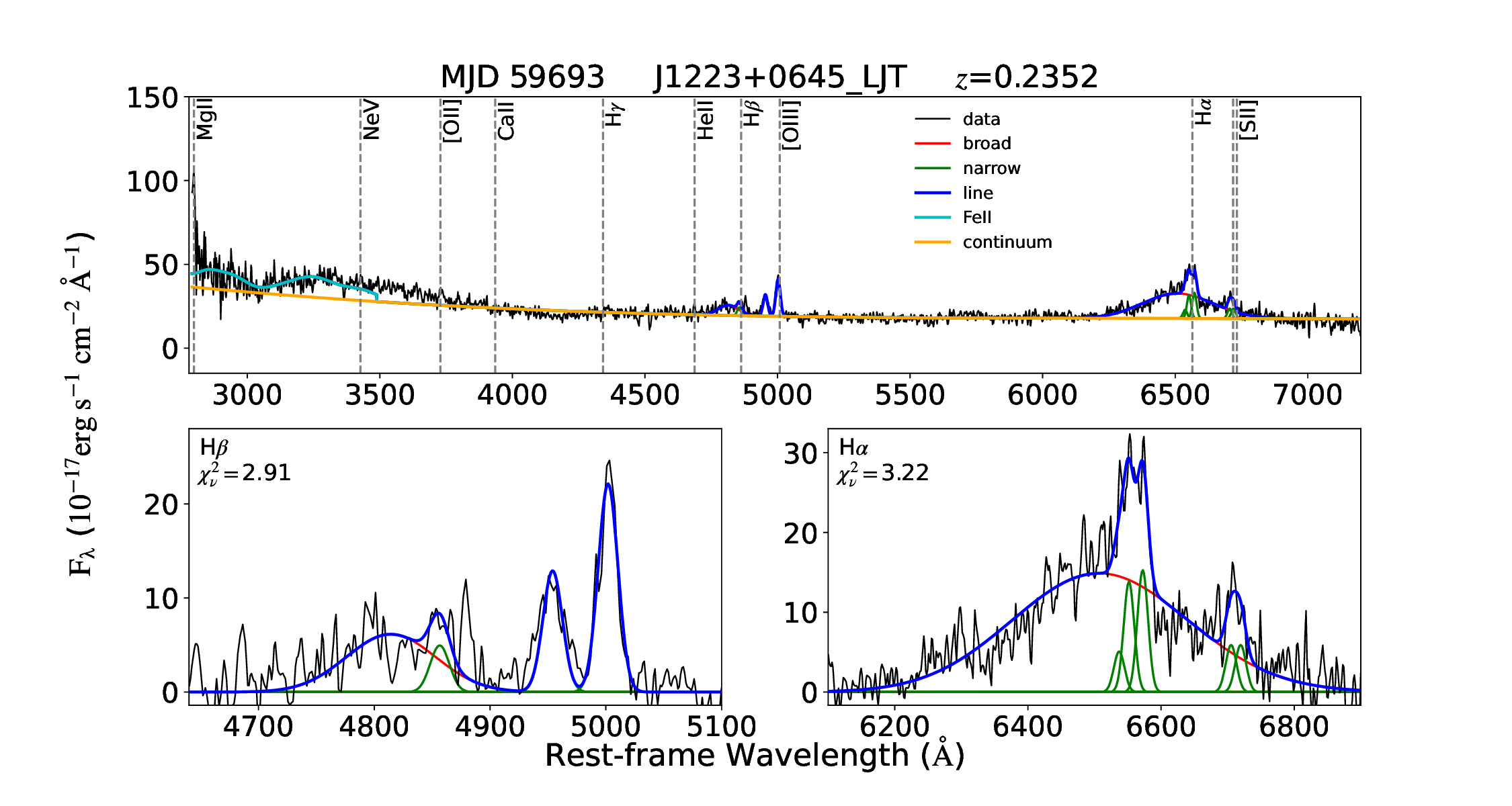}
	\caption{Spectral fitting for SDSS and LJT spectra of J1223+0645.}
	\label{fig:sf3}
	\end{figure*}

	\begin{figure*}
			\includegraphics[width=0.86\linewidth]{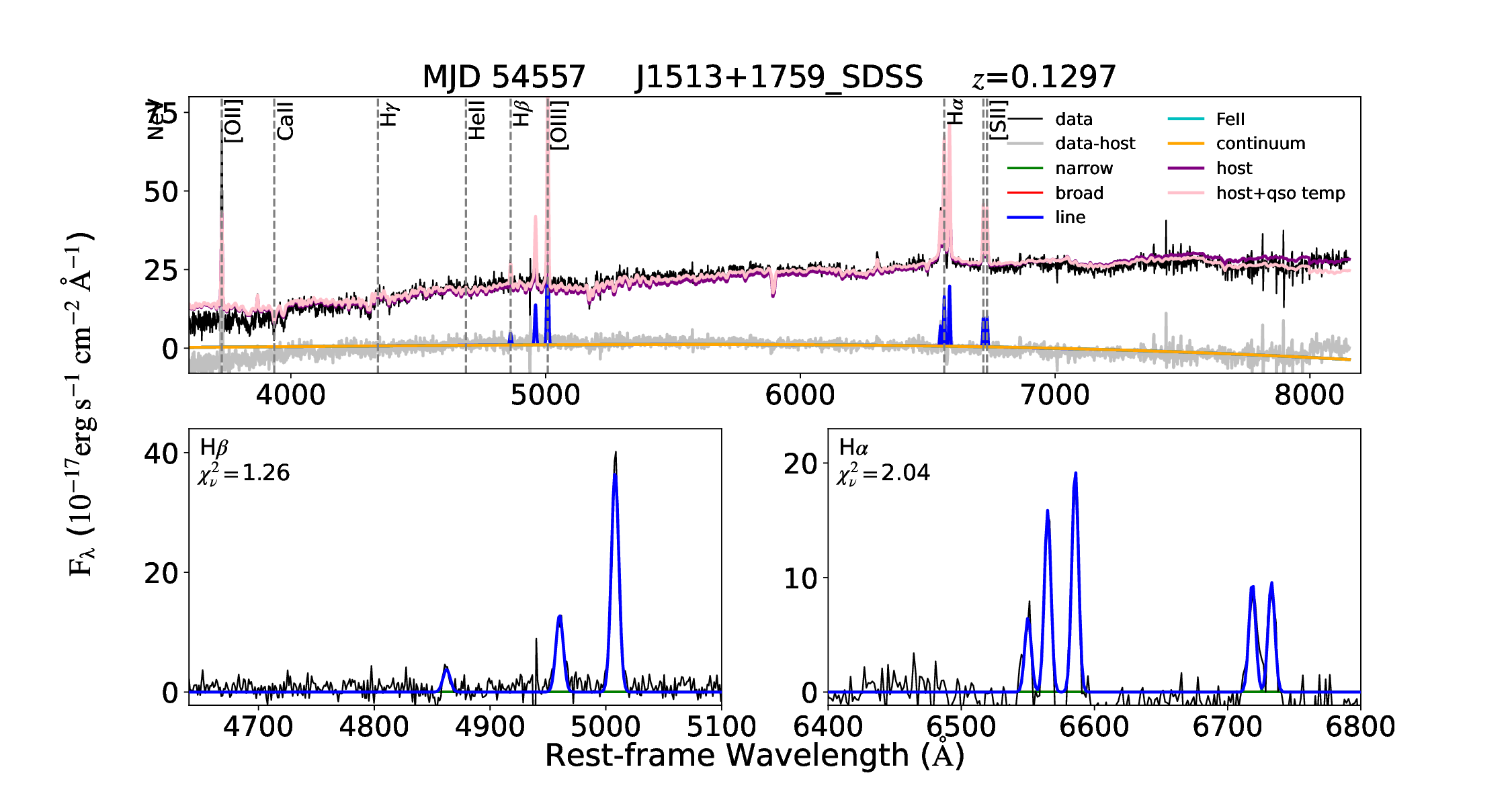}
			\includegraphics[width=0.86\linewidth]{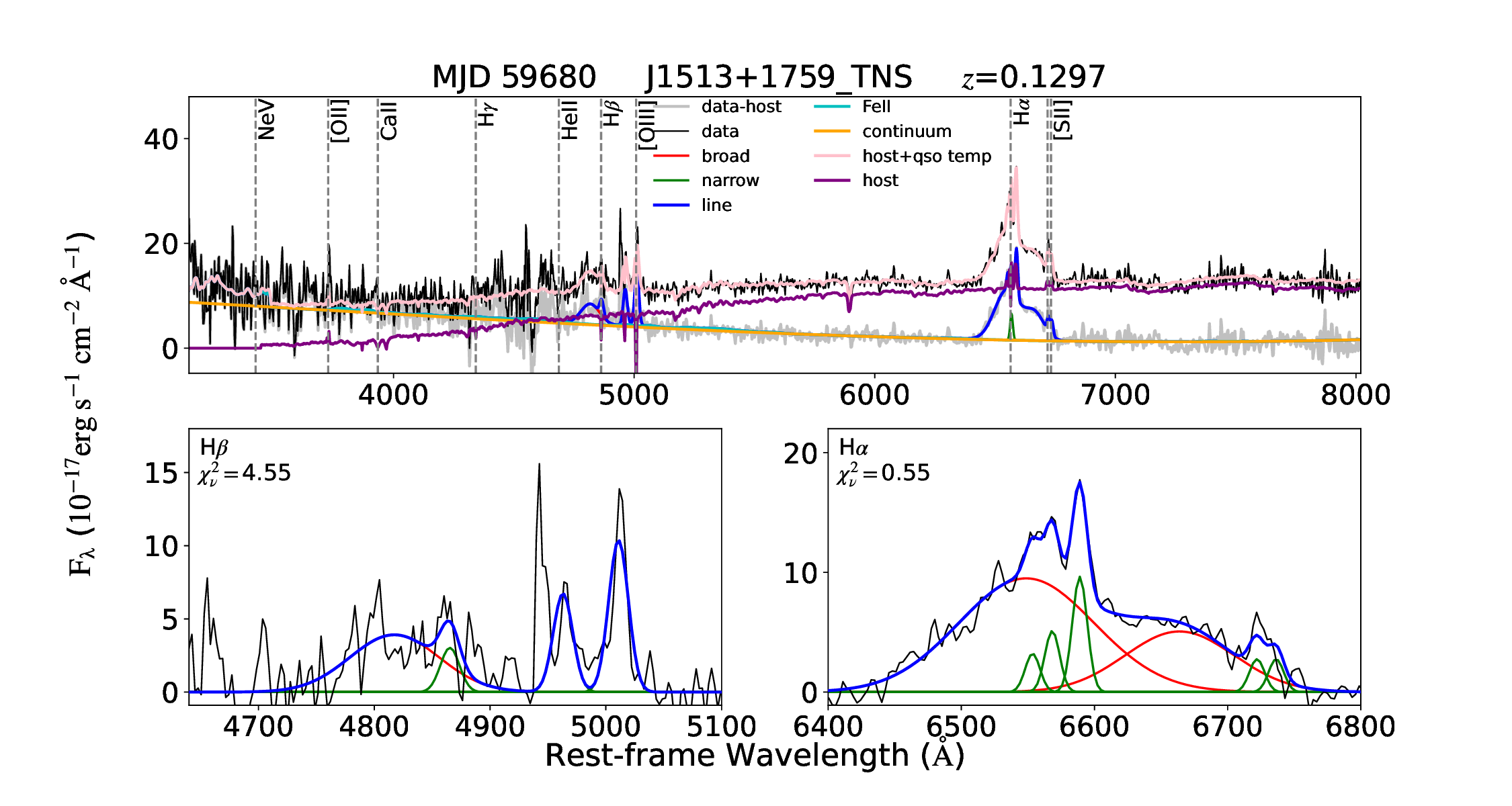}
			\includegraphics[width=0.86\linewidth]{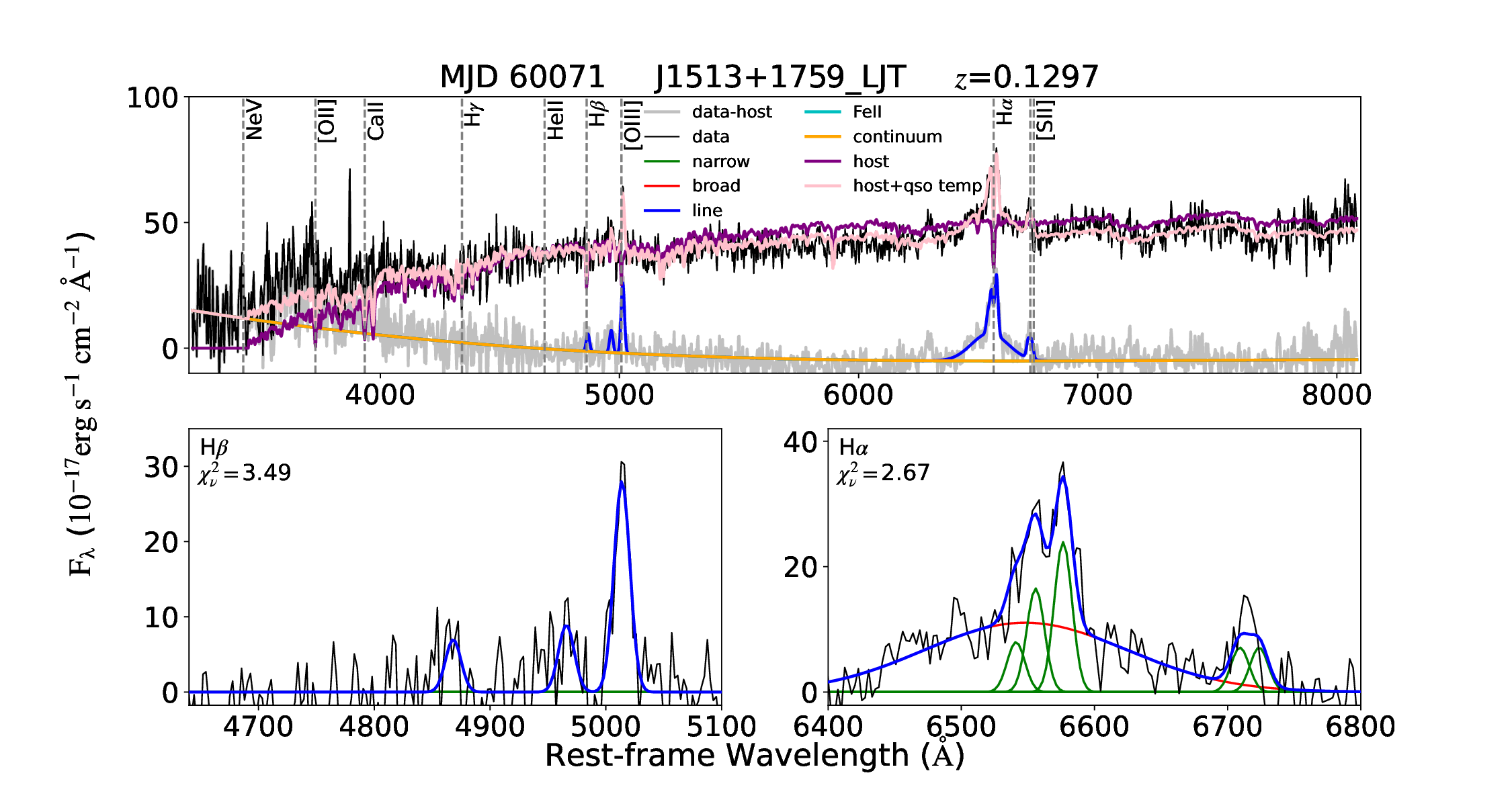}
	\caption{Spectral fitting for SDSS, TNS, and LJT spectra of J1513+1759.}
	\label{fig:sf4}

	\end{figure*}

\clearpage

\section{Magnitude color diagrams for three of the four AGN}
\label{sec:cm}

	\begin{figure}
		\includegraphics[width=0.86\linewidth]{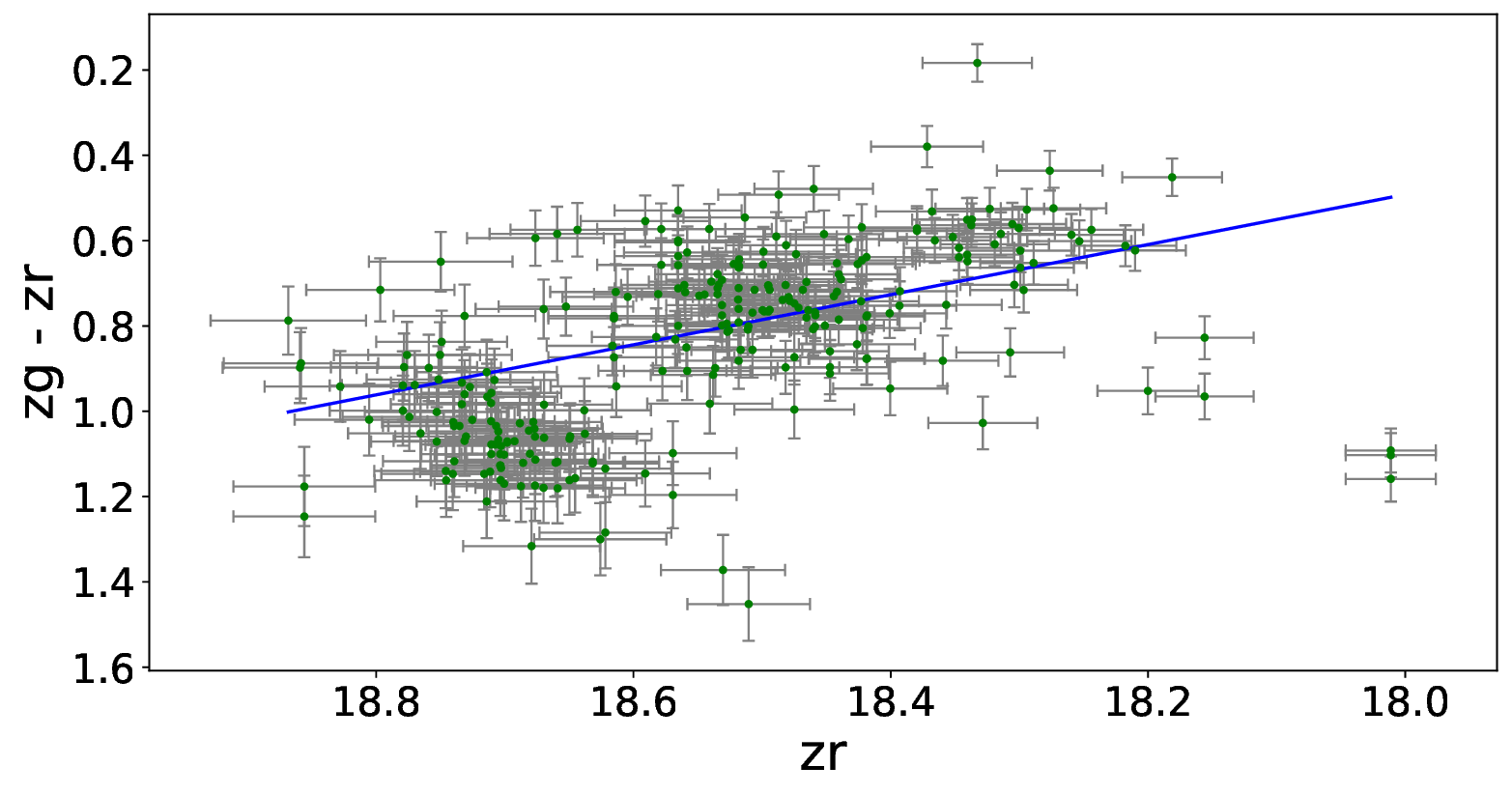}
                        \includegraphics[width=0.86\linewidth]{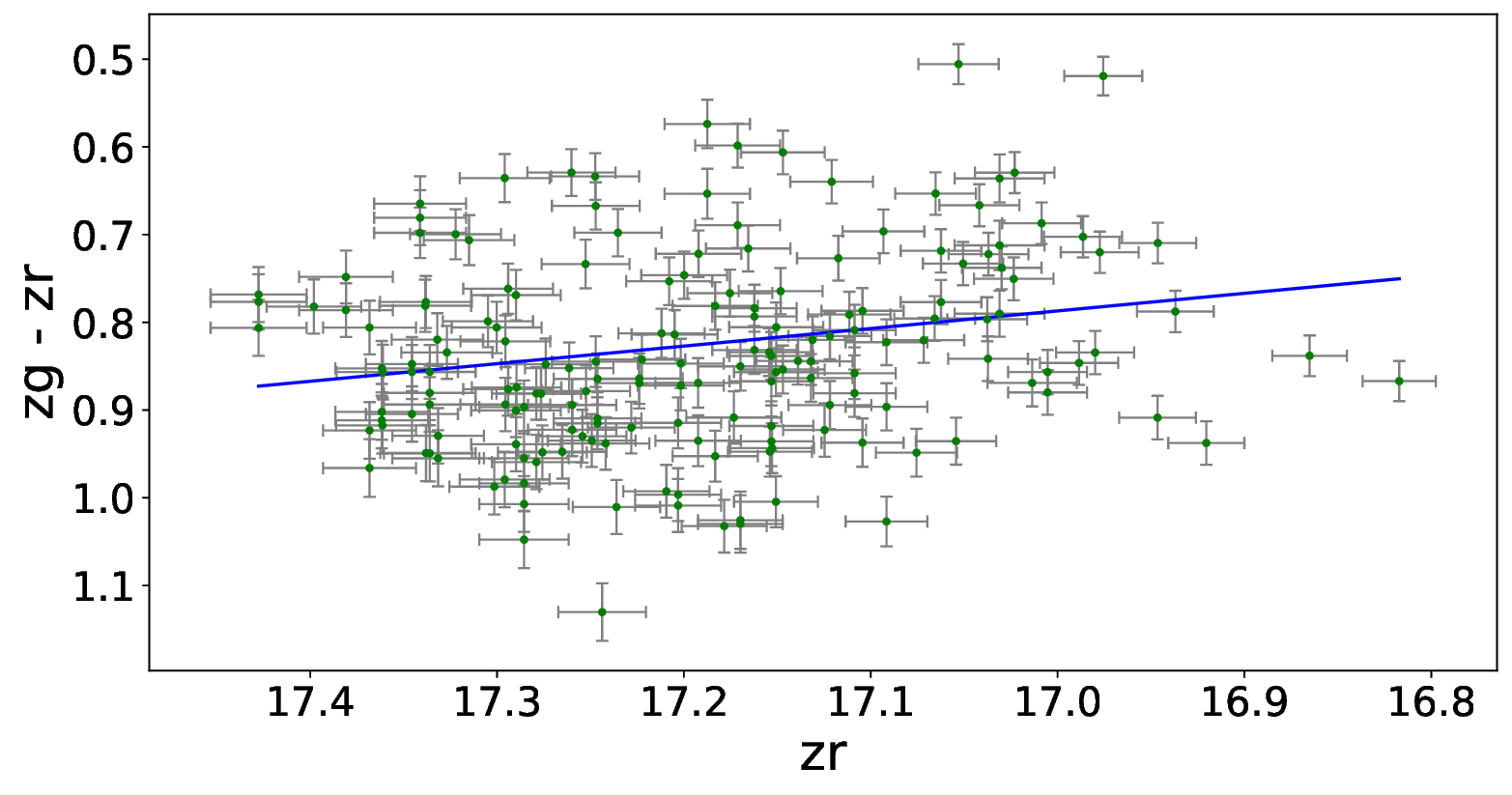}
                        \includegraphics[width=0.86\linewidth]{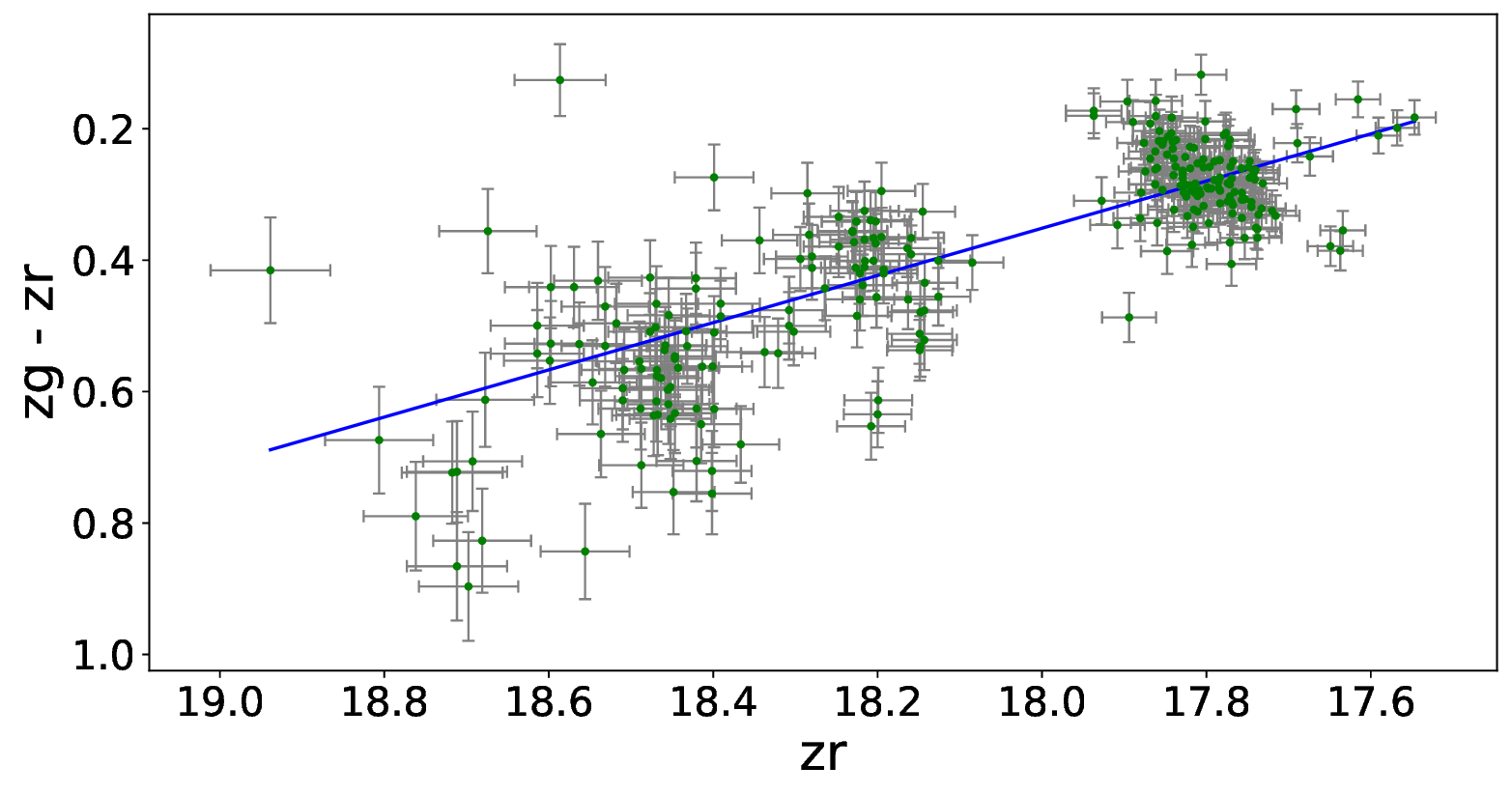}
		\caption{Magnitude color diagrams for J0113+0135 ({\it top}), 
		J1127+3546 ({\it middle}), and J1223+0645 ({\it bottom}). 
		The $zg - zr\propto k\times zr$ fitting lines to the data 
		points of the three sources are marked by a blue line in each 
		panel,
		where $k$ values are 0.59, 0.2, and 0.36, respectively.}
		\label{fig:cm}
	\end{figure}
\label{lastpage}
\end{document}